\documentclass[prb, twocolumn, showpacs, longbibliography, floatfix]{revtex4-1}
\usepackage{amsmath,amssymb,amsfonts,mathrsfs}
\usepackage{graphics, subfigure, color, bezier}
\usepackage{bm}
\usepackage{braket}
\usepackage{epstopdf}
\usepackage{dsfont}
\usepackage{ulem} 
\usepackage{color}

\usepackage[breaklinks]{hyperref}
\hypersetup{
   colorlinks,
   citecolor=blue,
   filecolor=blue,
   linkcolor=blue,
   urlcolor=blue
}

\usepackage{ifthen}		    

\usepackage{epsfig}
\usepackage{graphics, subfigure, color}

\newcommand{\e}[1]{{}_{\text{#1}}}
	{\begin{pmatrix}}%
	{\end{pmatrix}}
\newcommand{\f}[2]{{
	\ifmmode
		\relax
	\else
		\message{^^JERREUR: ligne \number\inputlineno, la commande 'f' devrait 
			être en mode math^^J^^J}
	\fi
	\mathchoice%
		{\dfrac{#1}{#2}}
    	{\dfrac{#1}{#2}}
		{\frac{#1}{#2}}
		{\frac{#1}{#2}}
}}

\newcommand{\moyenne}[2][9]{%
	\ifthenelse{#1 = 0}{#2}{}%
	\ifthenelse{#1 = 1}{\langle#2\rangle}{}%
	\ifthenelse{#1 = 2}{\big\langle#2\big\rangle}{}%
	\ifthenelse{#1 = 3}{\Big\langle#2\Big\rangle}{}%
	\ifthenelse{#1 = 4}{\bigg\langle#2\bigg\rangle}{}%
	\ifthenelse{#1 = 5}{\Bigg\langle#2\Bigg\rangle}{}%
	\ifthenelse{#1 = 9}{\left\langle#2\right\rangle}{}%
}
\newcommand{\ir}{\text{i}\,}

\newcommand{\SUM}[2]{{\displaystyle\sum\limits_{#1}^{#2}}}
\newcommand{\marge}{\noindent}
\newcommand{\ex}[1]{\text{e}^{\,#1}\,}

\newcommand{\dd}{\mathrm{d}}
\newcommand{\paf}[2]{\left(\f{#1}{#2}\right)}
\newcommand{\limt}{\lim\limits}

\DeclareMathOperator{\ch}{ch\,}
\renewcommand{\cosh}{\ch}					

\renewcommand{\coth}{\mathop{\rm coth\,}\nolimits}
\DeclareMathOperator{\cotan}{cotan\,}

\newif\ifdraft
\draftfalse 

\begin{document}

\title{Double-periodic Josephson junctions in a quantum dissipative environment}

\author{Tom Morel}
\author{Christophe Mora}
\affiliation{Laboratoire de physique de l'\'Ecole Normale Sup\'erieure, PSL Research University, CNRS, Universit\'e Pierre et Marie Curie-Sorbonne Universit\'es, Universit\'e Paris Diderot-Sorbonne Paris Cit\'e, 24 rue Lhomond, 75231 Paris Cedex 05, France}

\begin{abstract}
  Embedded in an ohmic environment, the Josephson current peak can transfer part of its weight to finite voltage and the junction becomes resistive. The dissipative environment can even suppress the superconducting effect of the junction via a quantum phase transition occuring when the ohmic resistance $R_s$ exceeds the quantum resistance $R\e{q}=h/(2e)^2$. For a topological junction hosting Majorana bound states with a $4 \pi$ periodicity of the superconducting phase, the phase transition is shifted to $4 R\e{q}$. We consider a Josephson junction mixing the $2 \pi$ and $4 \pi$ periodicities shunted by a resistor, with a resistance between $R_q$ and $4 R_q$. Starting with a quantum circuit model, we derive the non-monotonic temperature dependence of its differential resistance resulting from the competition between the two periodicities; the  $4 \pi$ periodicity dominating at the lowest temperatures. The non-monotonic behaviour is first revealed by straightforward perturbation theory and then substantiated by a fermionization to exactly solvable models when $R_s=2R\e{q}$: the model is mapped onto a  helical wire coupled to a topological superconductor when the Josephson energy is small and to the Emery-Kivelson line of the two-channel Kondo model in the opposite case.

\end{abstract}

\maketitle 

\section{Introduction}

The tunneling of Cooper pairs in the Josephson effect can be reduced and even suppressed by a shunting resistance $R\e{s}$. The resistor acts as an ohmic dissipative environment which controls the quantum fluctuations of the superconducting phase in the Josephson junction~\cite{ingold}. A renormalization group (RG) analysis predicts a quantum phase transition between a superconducting and an insulating state in a single Josephson junction~\cite{schmid,shon_zaikin,fisher_zwerger, zwerger, tewari, herrero, kimura, kohler, werner, lukyanov}. The location of the quantum phase transition is determined solely by the dimensionless dissipation strength $\alpha=R\e{q}/R\e{s}$ where $R\e{q}=h/(2e)^2$ is a quantum of resistance. When $\alpha>1$,
quantum fluctuations of the phase are suppressed by dissipation and the junction is superconducting. Conversely, for $\alpha<1$, the dissipation is strong enough to 
destroy the Josephson current even at zero temperature. Several aspects of this transition have been observed experimentally~\cite{yagi,penttila, kuzmin} in superconducting junctions shunted by metallic resistors.

The model describing the quantum phase transition is well-established and understood. It can be mapped onto the problem of quantum Brownian motion in a periodic potential which has been studied in detail~\cite{fisher_zwerger, aslangul, korshunov}. It is also equivalent to the one-dimensional boundary Sine-Gordon model~\cite{Zamolodchikov} which describes in particular an impurity in a Luttinger liquid~\cite{kane_fisher, giamarchi,fabrizio}, such as a defect in an interacting nanowire or a point contact in a fractional quantum Hall state~\cite{saleur-fendley}. More generally, the quantum phase transition and environment fluctuations have a strong impact on the whole current-voltage characteristics of the junction at energies well below the gap~\cite{ingold, grabert1999, corlevi, didier}.

The past years have witnessed a tremendous interest for the fractional Josephson effect in junctions hosting Majorana bound states~\cite{kitaev,beenakker}. Majorana excitations exhibit a topological protection against small perturbation and, as such, are believed to be building blocks for fault-tolerant quantum computation~\cite{ioffe,read,nayak} via their braiding~\cite{malciu}. The fractional Josephson effect involves a $4\pi$ periodicity of the current as function of the superconducting phase in contrast with the usual $2\pi$ periodicity. It has been tested experimentally in semiconducting nanowires and topological junctions via the absence of odd Shapiro steps under radio-frequency irradiation~\cite{rokhinson, wiedenmann, bocquillon}. Physically, the $4\pi$ periodicity is in fact associated with coherent single-electron tunneling at zero energy.

Topological junctions most probably combine Josephson energy terms with $2\pi$ and $4\pi$ periodicity. These multiple periodicities are nevertheless not uncommon since non-sinusoidal Josephson junctions, for instance in atomic point contacts~\cite{golubev,janvier}, already involve different harmonics associated with the presence of Andreev levels. At zero energy, an Andreev state produces a $4\pi$-periodic Josephson effect similar to the topological case. The presence of a strong Kondo impurity in the junction has been argued to pin the Andreev level to zero energy~\cite{zazuno}, thereby achieving a robust fractional Josephson effect. Moreover, there exist other means to realize different periodicities, including hybrid junctions involving superconducting and ferromagnetic layers which have been theoretically predicted to exhibit a controllable Josephson periodicity~\cite{ouassou}. Another proposal is a specific arrangement of four Josephson junctions  with a $\pi$ periodicity called the Josephson rhombus and also appearing in certain Josephson arrays~\cite{Ulrich_rhombus, gladchenko, bell, doucot, ioffe}. 

In this paper, we study a Josephson junction having the two periodicities $2 \pi$ and $4 \pi$ shunted by an ohmic environment. Whereas we use here a full quantum treatment, the classical limit of this model has been investigated in the framework of the  resistively capacitively shunted junction (RCSJ) model with the purpose of describing Shapiro steps~\cite{dominguez, pico, hangli,sau}. For a pristine topological Josephson junction with $4 \pi$ periodicity, a renormalization group analysis~\cite{ribeiro} shows that the superconductor-insulator quantum phase transition is just shifted to the critical value $ \alpha=R\e{q}/R\e{s} = 1/4$, four times smaller than for conventional Josephson junctions. This critical condition also reads $R_{q2}/R\e{s} = 1$, with $R_{q2} = h/e^2$. It is then easily understood by noting that single-electron tunneling occurs through Majorana bound states in topological Josephson junctions in contrast to Cooper pair tunneling in conventional Josephson junctions. 

\begin{figure}
\begin{center}
\includegraphics[width =0.5\columnwidth]{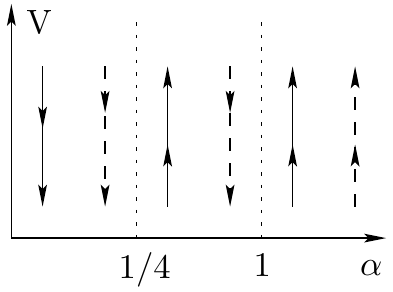}
\caption{Renormalization group scaling flows of the dissipative Josephson junction.  $V$ represents the strength of the potential term ($E_J$ or $E_M$). The full lines represent the $E_M$-term and the dashed lines the $E_J$-term. For $\alpha>1/4$, $E_M$ is relevant. We therefore expect that the system is in an insulating state $(R=R\e{s})$ for $\alpha<1/4$ and in a superconducting state $(R=0)$ for $\alpha>1/4$.
\label{fig_RG}}
\end{center}
\end{figure}

In the presence of both periodicities, a competition emerges with the phase diagram shown in Fig.~\ref{fig_RG}. We focus in this work on the values of $\alpha$ between $1/4$ and $1$ where, (i) the topological Josephson energy $E_M$, corresponding to single-electron tunneling, is relevant while (ii) the standard Josephson energy $E_J$, describing Cooper pair tunneling, is irrelevant but commensurate with the topological term. No intermediate fixed point can emerge from this competition since there are only two admissible infrared fixed points for the corresponding conformal field theory~\cite{CFT_affleck}, representing the superconducting and insulating states. Nevertheless, the two terms can dominate different energy regimes, $E_M$ being always the dominant effect at sufficiently low energy. We study the interplay of the two Josephson terms and the Coulomb interaction at arbitrary temperatures by using a combination of perturbative techniques and mappings to exactly solvable models for $\alpha = 1/2$. We compute the resistance of the whole system - Josephson junction and ohmic environment - as a function of temperature and exhibit non-monotonic behaviours for different regimes of Josephson and charging energies.

This article is organized as follows: in Sec. II we use a quantum circuit description of the system and show that the dissipative term and charging energy can be absorbed in the Josephson tunneling to recover the usual Sine-Gordon action~\cite{shon_zaikin,kane_fisher}.
The rest of the paper is devoted to the computation of the zero-bias differential resistance of the circuit at arbitrary temperature. In Sec. III, we identify the different low temperature regimes using renormalization group arguments. We then derive the resistance within linear response theory using perturbation theory and an infinite resummation based on refermionization at $\alpha=1/2$, thereby showing the non-monotonic temperature dependence. In Sec. IV, we treat with a tight-binding approach the limit of a deep Josephson periodic potential landscape with the Josephson energy much larger than the charging energy. A Bloch band description is combined with a refermionization procedure valid at $\alpha=1/2$ to derive a mapping to the Emery-Kivelson model of the two-channel Kondo problem. The topological Josephson energy $E_M$ acts as an effective magnetic field driving the system to a superconducting phase. We obtain an  analytical form for the resistance as function of temperature which qualitatively agrees with the shape derived in the opposite regime of small Josephson energies. We conclude in Sec. V.


\section{Circuit theory}

\subsection{Model}

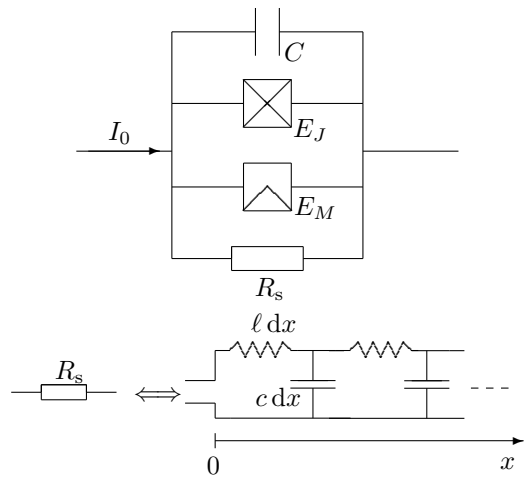
\begin{figure}
\begin{center}
\setlength{\unitlength}{3947sp}%
\begingroup\makeatletter\ifx\SetFigFont\undefined%
\gdef\SetFigFont#1#2#3#4#5{%
  \reset@font\fontsize{#1}{#2pt}%
  \fontfamily{#3}\fontseries{#4}\fontshape{#5}%
  \selectfont}%
\fi\endgroup%
\begin{picture}(2424,1888)(5839,-3737)
\put(6050,-2691){\makebox(0,0)[lb]{\smash{{\SetFigFont{10}{12.0}{\familydefault}{\mddefault}{\updefault}{\color[rgb]{0,0,0}$I_0$}%
}}}}
\thinlines
{\color[rgb]{0,0,0}\put(6901,-2311){\line( 1,-1){300}}
}%
{\color[rgb]{0,0,0}\put(6901,-2611){\line( 1, 1){300}}
}%
{\color[rgb]{0,0,0}\put(6901,-2461){\line(-1, 0){450}}
}%
{\color[rgb]{0,0,0}\put(7201,-2461){\line( 1, 0){450}}
}%
{\color[rgb]{0,0,0}\put(7276,-3436){\line( 1, 0){375}}
}%
{\color[rgb]{0,0,0}\put(6451,-2986){\line( 1, 0){450}}
\put(6901,-2986){\line( 0, 1){150}}
\put(6901,-2836){\line( 1, 0){300}}
\put(7201,-2836){\line( 0,-1){300}}
\put(7201,-3136){\line(-1, 0){300}}
\put(6901,-3136){\line( 0, 1){150}}
}%
{\color[rgb]{0,0,0}\multiput(6901,-3136)(6.00000,6.00000){26}{\makebox(1.6667,11.6667){\tiny.}}
\multiput(7051,-2986)(6.00000,-6.00000){26}{\makebox(1.6667,11.6667){\tiny.}}
}%
{\color[rgb]{0,0,0}\put(7201,-2986){\line( 1, 0){450}}
\put(7651,-2986){\line( 0, 1){525}}
}%
{\color[rgb]{0,0,0}\put(6451,-2461){\line( 0,-1){525}}
}%
{\color[rgb]{0,0,0}\put(6451,-2986){\line( 0,-1){450}}
}%
{\color[rgb]{0,0,0}\put(7651,-2986){\line( 0,-1){450}}
}%
{\color[rgb]{0,0,0}\put(7126,-2011){\line( 1, 0){525}}
}%
{\color[rgb]{0,0,0}\put(7126,-1861){\line( 0,-1){300}}
}%
{\color[rgb]{0,0,0}\put(6976,-1861){\line( 0,-1){300}}
}%
{\color[rgb]{0,0,0}\put(6451,-2461){\line( 0, 1){450}}
\put(6451,-2011){\line( 1, 0){525}}
}%
{\color[rgb]{0,0,0}\put(7651,-2011){\line( 0,-1){450}}
}%
{\color[rgb]{0,0,0}\put(7651,-2761){\line( 1, 0){600}}
}%
{\color[rgb]{0,0,0}\put(5926,-2761){\vector( 1, 0){450}}
}%
{\color[rgb]{0,0,0}\put(5851,-2761){\line( 1, 0){600}}
}%
{\color[rgb]{0,0,0}\put(6451,-3436){\line( 1, 0){375}}
\put(6826,-3436){\line( 0, 1){ 75}}
\put(6826,-3361){\line( 1, 0){450}}
\put(7276,-3361){\line( 0,-1){150}}
\put(7276,-3511){\line(-1, 0){450}}
\put(6826,-3511){\line( 0, 1){ 75}}
\put(6826,-3436){\line( 0, 1){ 75}}
}%
\put(7156,-2196){\makebox(0,0)[lb]{\smash{{\SetFigFont{10}{12.0}{\familydefault}{\mddefault}{\updefault}{\color[rgb]{0,0,0}$C$}%
}}}}
\put(7213,-2673){\makebox(0,0)[lb]{\smash{{\SetFigFont{10}{12.0}{\familydefault}{\mddefault}{\updefault}{\color[rgb]{0,0,0}$E_J$}%
}}}}
\put(7219,-3161){\makebox(0,0)[lb]{\smash{{\SetFigFont{10}{12.0}{\familydefault}{\mddefault}{\updefault}{\color[rgb]{0,0,0}$E_M$}%
}}}}
\put(6972,-3673){\makebox(0,0)[lb]{\smash{{\SetFigFont{10}{12.0}{\familydefault}{\mddefault}{\updefault}{\color[rgb]{0,0,0}$R_\text{s}$}%
}}}}
{\color[rgb]{0,0,0}\put(6901,-2311){\line( 0,-1){300}}
\put(6901,-2611){\line( 1, 0){300}}
\put(7201,-2611){\line( 0, 1){300}}
\put(7201,-2311){\line(-1, 0){300}}
}%
\end{picture}%
\setlength{\unitlength}{4144sp}%
\begingroup\makeatletter\ifx\SetFigFont\undefined%
\gdef\SetFigFont#1#2#3#4#5{%
  \reset@font\fontsize{#1}{#2pt}%
  \fontfamily{#3}\fontseries{#4}\fontshape{#5}%
  \selectfont}%
\fi\endgroup%
\begin{picture}(3089,1045)(1469,-2658)
\put(2943,-2191){\makebox(0,0)[lb]{\smash{{\SetFigFont{10}{12.0}{\familydefault}{\mddefault}{\updefault}{\color[rgb]{0,0,0}$c\,\text{d} x$}%
}}}}
\thinlines
{\color[rgb]{0,0,0}\put(2521,-2176){\line( 1, 0){180}}
\put(2701,-2176){\line( 0,-1){ 90}}
}%
{\color[rgb]{0,0,0}\put(2701,-2401){\vector( 1, 0){1845}}
}%
{\color[rgb]{0,0,0}\put(2701,-2356){\line( 0,-1){ 90}}
}%
{\color[rgb]{0,0,0}\put(4231,-2086){\line( 1, 0){ 45}}
}%
{\color[rgb]{0,0,0}\put(4321,-2086){\line( 1, 0){ 45}}
}%
{\color[rgb]{0,0,0}\put(4411,-2086){\line( 1, 0){ 45}}
}%
{\color[rgb]{0,0,0}\put(1481,-2111){\line( 1, 0){180}}
\put(1661,-2111){\line( 0, 1){ 45}}
\put(1661,-2066){\line( 1, 0){270}}
\put(1931,-2066){\line( 0,-1){ 90}}
\put(1931,-2156){\line(-1, 0){270}}
\put(1661,-2156){\line( 0, 1){ 45}}
\put(1661,-2111){\line( 0, 1){ 45}}
\put(1661,-2066){\line( 1, 0){270}}
\put(1931,-2066){\line( 0,-1){ 45}}
\put(1931,-2111){\line( 1, 0){180}}
}%
{\color[rgb]{0,0,0}\put(3196,-1861){\line( 1, 0){225}}
}%
{\color[rgb]{0,0,0}\put(3286,-1861){\line( 0,-1){180}}
\put(3286,-2041){\line(-1, 0){ 90}}
}%
{\color[rgb]{0,0,0}\put(3286,-2041){\line( 1, 0){ 90}}
}%
{\color[rgb]{0,0,0}\put(3151,-2041){\line( 1, 0){270}}
}%
{\color[rgb]{0,0,0}\put(3151,-2086){\line( 1, 0){270}}
}%
{\color[rgb]{0,0,0}\put(3286,-2086){\line( 0,-1){180}}
}%
{\color[rgb]{0,0,0}\put(2791,-2266){\line( 1, 0){720}}
}%
{\color[rgb]{0,0,0}\put(3421,-1861){\line( 1, 0){ 90}}
}%
{\color[rgb]{0,0,0}\put(2701,-1861){\line( 1, 0){ 90}}
}%
{\color[rgb]{0,0,0}\put(2701,-2266){\line( 1, 0){ 90}}
}%
{\color[rgb]{0,0,0}\put(3877,-1862){\line( 1, 0){225}}
}%
{\color[rgb]{0,0,0}\put(3967,-1862){\line( 0,-1){180}}
\put(3967,-2042){\line(-1, 0){ 90}}
}%
{\color[rgb]{0,0,0}\put(3967,-2042){\line( 1, 0){ 90}}
}%
{\color[rgb]{0,0,0}\put(3832,-2042){\line( 1, 0){270}}
}%
{\color[rgb]{0,0,0}\put(3832,-2087){\line( 1, 0){270}}
}%
{\color[rgb]{0,0,0}\put(3967,-2087){\line( 0,-1){180}}
}%
{\color[rgb]{0,0,0}\put(3472,-2267){\line( 1, 0){720}}
}%
{\color[rgb]{0,0,0}\put(4102,-1862){\line( 1, 0){ 90}}
}%
{\color[rgb]{0,0,0}\put(3382,-1862){\line( 1, 0){ 90}}
}%
{\color[rgb]{0,0,0}\put(3382,-2267){\line( 1, 0){ 90}}
}%
{\color[rgb]{0,0,0}\multiput(2791,-1861)(5.62500,5.62500){9}{\makebox(1.5875,11.1125){\tiny.}}
\multiput(2836,-1816)(3.75000,-7.50000){13}{\makebox(1.5875,11.1125){\tiny.}}
\multiput(2881,-1906)(3.75000,7.50000){13}{\makebox(1.5875,11.1125){\tiny.}}
\multiput(2926,-1816)(3.75000,-7.50000){13}{\makebox(1.5875,11.1125){\tiny.}}
\multiput(2971,-1906)(3.75000,7.50000){13}{\makebox(1.5875,11.1125){\tiny.}}
\multiput(3016,-1816)(3.75000,-7.50000){13}{\makebox(1.5875,11.1125){\tiny.}}
\multiput(3061,-1906)(3.75000,7.50000){13}{\makebox(1.5875,11.1125){\tiny.}}
\multiput(3106,-1816)(5.62500,-5.62500){9}{\makebox(1.5875,11.1125){\tiny.}}
\put(3151,-1861){\line( 1, 0){ 45}}
}%
{\color[rgb]{0,0,0}\multiput(3511,-1861)(5.62500,5.62500){9}{\makebox(1.5875,11.1125){\tiny.}}
\multiput(3556,-1816)(3.75000,-7.50000){13}{\makebox(1.5875,11.1125){\tiny.}}
\multiput(3601,-1906)(3.75000,7.50000){13}{\makebox(1.5875,11.1125){\tiny.}}
\multiput(3646,-1816)(3.75000,-7.50000){13}{\makebox(1.5875,11.1125){\tiny.}}
\multiput(3691,-1906)(3.75000,7.50000){13}{\makebox(1.5875,11.1125){\tiny.}}
\multiput(3736,-1816)(3.75000,-7.50000){13}{\makebox(1.5875,11.1125){\tiny.}}
\multiput(3781,-1906)(3.75000,7.50000){13}{\makebox(1.5875,11.1125){\tiny.}}
\multiput(3826,-1816)(5.62500,-5.62500){9}{\makebox(1.5875,11.1125){\tiny.}}
\put(3871,-1861){\line( 1, 0){ 45}}
}%
\put(2651,-2603){\makebox(0,0)[lb]{\smash{{\SetFigFont{10}{12.0}{\familydefault}{\mddefault}{\updefault}{\color[rgb]{0,0,0}$0$}%
}}}}
\put(4406,-2571){\makebox(0,0)[lb]{\smash{{\SetFigFont{10}{12.0}{\familydefault}{\mddefault}{\updefault}{\color[rgb]{0,0,0}$x$}%
}}}}
\put(2171,-2162){\makebox(0,0)[lb]{\smash{{\SetFigFont{10}{12.0}{\familydefault}{\mddefault}{\updefault}{\color[rgb]{0,0,0}$\iff$}%
}}}}
\put(1741,-2031){\makebox(0,0)[lb]{\smash{{\SetFigFont{10}{12.0}{\familydefault}{\mddefault}{\updefault}{\color[rgb]{0,0,0}$R_\text{s}$}%
}}}}
\put(2917,-1748){\makebox(0,0)[lb]{\smash{{\SetFigFont{10}{12.0}{\familydefault}{\mddefault}{\updefault}{\color[rgb]{0,0,0}$\ell\,\text{d} x$}%
}}}}
{\color[rgb]{0,0,0}\put(2701,-1861){\line( 0,-1){180}}
\put(2701,-2041){\line(-1, 0){180}}
}%
\end{picture}%
  \caption{(a) Schematic representation of a resistively and capacitively shunted Josephson junction combining $2\pi$ and $4\pi$-periodic contributions. (b) Sketch of the distributed LC line circuit representing the shunt resistor $R\e{s}$. The dissipative environment is described as a semi-infinite transmission line with lineic capacitance $c$ and lineic inductance $\ell$. The correspondance between the two representations gives $R\e{s}=\sqrt{\ell/c}$.
\label{fig_system}}
\end{center}
\end{figure}

Instead of starting from an abstract Caldeira-Leggett form, we derive the relevant Hamiltonian
from quantum circuit theory~\cite{michel,leppakangas}. We consider the quantum device
depicted in Fig.~\ref{fig_system} composed of three
parallel elements: a superconducting junction with a 
Josephson energy $E_J$, a second topological junction with a Josephson energy $E_M$,
a capacitance $C$ and a resistor 
$R\e{s}$. The whole apparatus is biased by a dc-current $I_0$.
The fractional Josephson junction allows for coherent single-electron tunneling \textit{i.e} a $4\pi$-periodicity of the phase. We neglect in our analysis the quasiparticle excitations above the superconducting gap and we use $\hbar=k_B=1$ for simplicity.

In the charge representation, the Hamiltonian of the system is
\begin{multline}\label{hamilto}
H=\f{1}{2C}\,\bigg(2e\,\hat{N}+\hat{Q}+\displaystyle\int_{-\infty}^{t}\dd t'\,I_0(t')\bigg)^2+H_{R\e{s}}\\
-\f{E_J}{2}\,
\SUM{n}{} \Big(|n\rangle\,\langle n+1| + \text{h.c}\Big)
-\f{E_M}{2}\,
\SUM{n}{} \Big(|n\rangle\,\langle n+\f{1}{2}| + \text{h.c}\Big)
\end{multline}
The first term is the energy stored in the capacitance where the 
charge $2e\,\hat{N}$ across the Josephson junctions is added to the 
charge $\hat{Q}$ brought by the resistor and the charge integrated from the current source $I_0$.
The third term corresponds 
to the Josephson tunneling between states with consecutive Cooper pair charge
numbers, $\hat{N}\,|n\rangle=n\,|n\rangle$. The fourth term describes the tunneling of
electrons through the topological Majorana fermions, implying that the charge operator $\hat{N}$
takes half-integer values, such that the corresponding phase
operator
\begin{equation}\label{phaseoperator}
\ex{\ir\hat{\varphi}/2}=\SUM{n}{} |n\rangle\langle n+1/2|,
\end{equation}
with  $[\hat{\varphi},\hat{N}]=\ir$, is defined on a circle of size $4 \pi$.
An electron is thus seen as half of a Cooper pair.
$H_{R\e{s}}$ models
the resistor in terms of an semi-infinite one-dimensional
 transmission line, \textit{i.e} as a 
collection of harmonic oscillators with lineic inductance $\ell$ and 
capacitance $c$~\cite{thesis}. The Hamiltonian $H_{R\e{s}}$ is
\begin{equation}
  H_{R\e{s}}[\{\hat{\phi}\},\{\hat{q}\}] = \int_0^{+\infty} d x \left [ \frac{1}{2 \ell} \left ( \frac{\partial  \hat \phi (x)}{\partial x} \right)^2 + \frac{\hat q (x)^2}{2 c} \right]
\end{equation}
where $R\e{s}=\sqrt{\ell/c}$. The local flux $\hat{\phi}$ and the local 
charge $\hat{q}$ are conjugate variables and obey the canonical 
quantization 
\begin{equation}\label{cano-quant}
  [ \hat \phi (x), \hat q (x') ] = i  \delta(x-x'),
\end{equation}
with the additional constraint that $\hat{Q}$ and $\hat \phi (0)$ are conjugate operators
\begin{equation}\label{cano-quant2}
[ \hat \phi (0), \hat Q ] = \ir
\end{equation}

\subsection{Unitary transformation}\label{sec-unitary}

Before acting on the Hamiltonian~\eqref{hamilto}, we note that the Hamiltonian
$\hat{Q}^2/2 C + H_R$ can be diagonalized by the following mode expansion
\begin{subequations}
\begin{align}
\label{phi-field}
  \hat \phi (x) & = \sqrt{\frac{R\e{s}}{4 \pi}} \int_0^{+\infty} \frac{d \omega}{\sqrt{\omega}} \left[ a_{\rm in,\omega} \ex{-\ir k x} 
  + a_{\rm out,\omega} \ex{\ir k x} + {\rm h.c.} \right] \\ 
  \label{q-field}
  \hat q (x) & = \frac{c}{i}  \,  \sqrt{\frac{R\e{s}}{4 \pi}} \int_0^{+\infty} \sqrt{\omega} \, d \omega [ a_{\rm in,\omega} \ex{-\ir k x} 
 + a_{\rm out,\omega} \ex{\ir k x} - {\rm h.c.}]
\end{align}
\end{subequations}
with the  dispersion $\omega = v k$ and the velocity $v = 1/\sqrt{\ell c}$, and the boundary conditions
\begin{equation}
 a_{\rm out,\omega} = \frac{1+\ir\tau\e{s}\omega }{1-\ir\tau\e{s}\omega} \, a_{\rm in,\omega}, \qquad \hat Q = \frac{C}{c} \hat q(0),
\end{equation}
corresponding to the reflexion of microwaves by the capacitor. The commutation relations Eqs.~\eqref{cano-quant} and~\eqref{cano-quant2} are recovered from the canonical quantization
\begin{equation}
 [a_{ \rm in,\omega}, a_{\rm in,\omega'}^\dagger  ] = \delta(\omega-\omega').
\end{equation}
This is a field theoretical description of a simple RC circuit with the time scale for discharge $\tau_s = R_s C$. Inserting this mode expansion, we find the diagonal form
\begin{equation}
  \frac{ \hat{Q}^2}{2 C} + H_{R\e{s}} =  \int_0^{+ \infty} \dd\omega \,  \omega  \, a_{\rm in,\omega}^\dagger a_{\rm in,\omega}.
\end{equation}
In order to disentangle the different variables, it is convenient to apply the time-dependent unitary transformation
 \begin{equation}\label{transformation}
 \hat{U} = \exp \left[ \ir \hat \phi (0) \left( 2  e \hat N + \int_{-\infty}^{t}\dd t'\,I_0(t') \right) \right]
 \end{equation}
 which essentially shifts the charge operator
 \begin{equation}
   \hat{U} \hat{Q} \hat{U}^\dagger =  \hat{Q} - 2  e \hat N - \int_{-\infty}^{t}\dd t'\,I_0(t'),
 \end{equation}
 and acts as a displacement operator for the propagating modes
 \begin{equation}\label{shift}
  \hat{U}  a_{\rm in,\omega} \hat{U}^\dagger =  a_{\rm in,\omega} -\ir \sqrt{\frac{R\e{s}}{\pi \omega}} \, \frac{2e\,\hat{N} +\int_{-\infty}^{t}\dd t'\,I_0(t')  }{1+\ir \tau\e{s} \omega}
\end{equation}
We note however that $\hat{U}$ leaves $ H_{R\e{s}} = \hat{U} H_{R\e{s}} \hat{U}^\dagger $ invariant.
The transformed Hamiltonian $\tilde{H}=\hat{U} \,H \,\hat{U}^\dagger + i \partial_t \hat{U} \hat{U}^\dagger$ assumes the simplified form
\begin{equation}\label{hamilto3}
  \begin{split}
  \tilde{H} =  &- \frac{E_J}{2} \sum_n \left( | n \rangle \langle n+1 | \ex{- 2 \ir e \hat \phi (0)} + \text{h.c} \right)\\
 & -\f{E_M}{2}\,
\SUM{n}{} \Big(|n\rangle\,\langle n+\f{1}{2}| \,\ex{-\ir e \hat \phi (0)}+ \text{h.c}\Big)
\\  & + \int_0^{+ \infty} \dd\omega \,  \omega  \, a_{\rm in,\omega}^\dagger a_{\rm in,\omega} - I_0(t) \, \hat{\phi} (0),
  \end{split}
\end{equation}
where the Cooper and single-electron tunneling terms are dressed by the dissipative phase functions
$e^{ \pm p \ir e \hat \phi (0)}$, with $p=1,2$, describing the RC environment.
Using Eq.\eqref{phaseoperator}, the Eq.\eqref{hamilto3} becomes 
 \begin{multline}\label{hamilto4}
  \tilde{H}=-E_J\,
  \cos\Big(\hat{\varphi}-2e\,\hat{\phi}(0)\Big)-E_M\,
    \cos\Big(\f{\hat{\varphi}}{2}-e\,\hat{\phi}(0)\Big)\\
    +\int_0^{+ \infty} \dd \omega\, \omega  \, a_{\rm in,\omega}^\dagger \,a_{\rm in,\omega}- I_0(t) \, \hat{\phi} (0).
\end{multline}
 At this point, $\hat N$ disappeared and the phase operator $\hat{\varphi}$ commutes with the Hamiltonian $\tilde{H}$.  $\hat{\varphi}$ is a constant of motion and it can be absorbed into  $\hat{\phi}(0)$, {\it i.e.} removed from Eq.~\eqref{hamilto4}. We emphasize that, although we made no approximation, the charge discreteness and the related phase compactness no longer play a role in Eq.~\eqref{hamilto4}.

 It is also be possible to formulate the euclidean action corresponding to the Hamiltonian~\eqref{hamilto4}, where all modes except $\phi_0 \equiv 2 e \hat{\phi} (0)$ are integrated,
\begin{multline}\label{action}
S=\f{1}{2\beta}\,\SUM{\ir\omega_n}{}\Big(\f{1}{8E_C}\,{\omega_n}^2+\f{\alpha}{2\pi}\,|\omega_n|\Big)\,|\phi_0(\ir\omega_n)|^2\\
-E_J\,\displaystyle\int_{0}^{\beta}\dd\tau\,\cos \phi_0(\tau) -E_M \,\displaystyle\int_{0}^{\beta}\dd\tau\,\cos\left( \frac{\phi_0(\tau)}{2} \right),
\end{multline}
for $I_0=0$. This expression recovers the standard action already used by many authors~\cite{shon_zaikin}  for 
$E_M=0$. We have introduced the charging energy $E_C = e^2/2 C$ and the dimensionless dissipative constant  $\alpha=R\e{q}/R\e{s}$ where $R\e{q}=h/(4e^2)$ is  the quantum resistance for Cooper pairs.

Hereinafter, we will use equivalently Eq.~\eqref{hamilto4} and Eq.~\eqref{action} as a starting point to derive the differential resistance of our model.

\section{Differential resistance in the Coulomb blockade regime}\label{sec-coulomb}


\subsection{Linear response theory} 

The effective resistance of the circuit is defined by the relation $V=R(T)\,I_0$
where $I_0$ is the  bias current and the voltage drop across the junction is
$$V=  \langle \partial_t \hat{\phi} (x=0) \rangle,$$
We use linear response theory to compute $R(T)$ by treating $- I_0(t) \, \hat{\phi} (0)$ in Eq.~\eqref{hamilto4} as a perturbation. The details in the imaginary time formalism are provided in appendix~\ref{appen-linear} where the expression 
 \begin{equation}{\label{resistance}}
 \f{R}{R\e{s}}=1+\f{2\pi}{\alpha}\,\limt_{\omega\to 0}
 \,\mathcal{R}\Big(\f{\ir}{\omega}\,\limt_{\ir\omega_n\to \omega+\ir 0^+}
 \,\displaystyle\int_{0}^{\beta}\,\ex{\ir\omega_n\tau}\langle\hat{f}(\tau)\hat{f}(0)\rangle\Big)
 \end{equation}
 is derived. $\omega_n = 2\pi \,n T$ denotes a Matsubara frequency,  $\mathcal{R}$ the real part, and $\beta=1/T$ the inverse temperature. We have also introduced the current-like operator
 \begin{equation}{\label{matsubara}}
 \hat{f}(\tau)=E_J\sin[2e\,\hat{\phi}(\tau)]+\f{E_M}{2}\,\sin [e\,\hat{\phi}(\tau)]
 \end{equation}
where we use the notation  $\hat{\phi}(\tau)=\hat{\phi}(x=0,\tau)$.
For $E_M=E_J=0$, we recover $R=R\e{s}$ as expected.

The computation of the resistance~\eqref{resistance} is based on the evaluation of the phase autocorrelation functions $\langle \ex{\ir p e\,\hat{\phi}(\tau)}\, \ex{-\ir pe\,\hat{\phi}(0)} \rangle$, with $p=1,2$. This can be done in perturbation theory in $E_J,E_M$, with the expression of the phase at $x=0$,
\begin{equation}\label{phi}
\hat{\phi}(0)=\sqrt{\f{R\e{s}}{4\pi}}\,\displaystyle\int_{0}^{+\infty}
\f{\dd\omega}{\sqrt{\omega}}\,\bigg[\f{2}{1-\ir\tau\e{s}\omega}\,
a\e{in,$\omega$}+ \text{h.c}\bigg],
\end{equation}
the thermal occupation
\begin{equation}
  \langle a\e{in,$\omega$}^\dagger a\e{in,$\omega'$} \rangle = f_B (\omega) \delta(\omega-\omega'),
\end{equation}
and the Bose factor $f_B (\omega) = (e^{\beta \omega}-1)^{-1}$. The leading order ($E_J,E_M=0$) is given at zero temperature and in real time by the expression familiar to the $P(E)$ theory~\cite{ingold, moskova,golubev, grabert, joyez}
\begin{equation}\label{correlator}
  \begin{split}
   &   \langle \ex{\ir p e\,\hat{\phi}(t)}\, \ex{-\ir pe\,\hat{\phi}(0)} \rangle = e^{J(t,p)}, \\[2mm]
  &  J(t,p)= \f{p^2}{2}\,\displaystyle\int_{-\infty}^{+\infty}\f{\dd\omega}{\omega} \f{{\rm Re} Z(\omega)}{R_q}
\left( e^{- i \omega t} -1 \right),
  \end{split}
\end{equation}
with the impedance of the RC environment $ Z(\omega) = (i \omega C+1/R_\text{s})^{-1}$.

Physically, the long-time asymptotics
\begin{equation}
  e^{J(t,p)} \sim \left( \frac{\tau_s}{t} \right)^{p^2/2 \alpha}
\end{equation}
measures how fast phase correlations decay in real time. A large $\alpha$ corresponds to a slow diffusion indicating a well-defined superconducting phase. The result is that the second term with the time integral in Eq.~\eqref{resistance} diverges as $\omega \to 0$ at zero temperature indicating a breakdown of perturbation theory and a flow towards zero resistance, {\it i.e.} a superconducting state with a Josephson current. In contrast, a small $\alpha$ gives a fast phase diffusion resulting in a vanishing second term in  Eq.~\eqref{resistance} for $\omega,T=0$,  {\it i.e.} an insulating state with $R = R_s$. Just by power counting, the threshold between these two states is found at $\alpha = p^2/4$, as recapitulated in Fig.\ref{fig_RG}.

\subsection{Perturbation theory for the resistance}

After setting the basis of the calculation in linear response theory, we compute the resistance at finite temperature from Eq.~\eqref{resistance} and perturbatively in $E_J,E_M\ll E_C$. The leading corrections to the fully shunted junction are derived in appendix~\ref{appen-perturbation} and  expressed as~\cite{fisher_zwerger}
\begin{multline}\label{perturbation}
\f{R(T)}{R\e{s}}=1-\f{{E_J}^2}{2\alpha T^2}\,
\displaystyle\int_{0}^{+\infty}\,\dd y \, \ex{ j_2(y,2 \alpha E_C/\pi^2 T)}\\-
\f{{E_M}^2}{8\alpha T^2}\,
\displaystyle\int_{0}^{+\infty}\,\dd y \,
 \ex{j_1(y,2 \alpha E_C/\pi^2 T)}
\end{multline}
with the functions
\begin{multline}\label{perturbation2}
j_p(y,a) = -\f{p^2}{2\alpha}\,\bigg\{\gamma+\Psi(a)+\ln 2 +\ln\ch\paf{y}{2}
\\
+\f{1}{2}\bigg[\f{1}{a}+\f{\pi\,\ex{-ay}}{\sin(\pi a)}\bigg]\\
-\f{\ex{-y}}{2}\bigg[\f{1}{1-a}\,{}_2F_1\Big(1,1-a,2-a,-\ex{-y}\Big)
\\
+\f{1}{1+a}\,{}_2F_1\Big(1,1+a,2+a,-\ex{-y}\Big)\bigg]
\Bigg\}
\end{multline}
$\gamma$ is the Euler's constant, $\Psi$ the 
logarithmic derivative of the Gamma function and ${}_2F_1$ 
an hypergeometric function.
The result~\eqref{perturbation} is only valid perturbatively {\it i.e.} when the second and third terms are much smaller than $1$. It can be further simplified at low temperatures $T \ll 2 \alpha E_C/\pi^2$, leading to 
\begin{multline}\label{low}
\f{R(T)}{R\e{s}}=1-K_J\,\paf{E_J}{E_C}^2\,\paf{T}{E_C}^{-2+2/\alpha}\\
-  K_M\,\paf{E_M}{E_C}^2\,\paf{T}{E_C}^{-2+1/(2\alpha)}
\end{multline}

\noindent with 

\begin{subequations}\label{coefficientK}
\begin{align}
K_J&=\f{\pi^{\frac{1}{2}+\frac{4}{\alpha}}\,\Gamma(\frac{1}{\alpha})}{(2\alpha)^{1+\frac{2}{\alpha}}\,(2\ex{\gamma})^{\frac{2}{\alpha}}\,\Gamma(\frac{1}{2}+\frac{1}{\alpha})}\\[3mm]
K_M &= \f{\pi^{\frac{1}{2}+\frac{1}{\alpha}}\,\Gamma(\frac{1}{4\alpha})}{4\,(2\alpha)^{1+\frac{1}{2\alpha}}\,(2\ex{\gamma})^{\frac{1}{2\alpha}}\,\Gamma(\frac{1}{2}+\frac{1}{4\alpha})}
\end{align}
\end{subequations}

\noindent with $\Gamma(x)$ is the Gamma function. We recover the results of 
the renormalisation group analysis that $E_J$ (resp.  $E_M$) scales down to zero as the temperature is lowered
for $\alpha < 1$ (resp. $\alpha < 1/4$) whereas it becomes increasingly large at low energy in the opposite case
$\alpha>1$  (resp. $\alpha > 1/4$).

We focus henceforth on the most interesting scenario where $\alpha$ is chosen between $1/4$ and $1$, such that $E_M$ is relevant and $E_J$ is irrelevant at low energy. There, a competition emerges between the two temperature corrections of Eq.~\eqref{low} with opposite limits. Since $E_M$ eventually dominates at sufficiently low energy, the competition is best discussed in the regime $E_M \ll E_J$. Differentiating $R(T)$ with respect to $T$, we find that the resistance reaches a (local) maximum for

\begin{equation}
  \f{T_m}{E_C} =\f{4 \ex{\gamma} \alpha}{\pi^2} \,\bigg[\f{(4\alpha-1)\,\tilde{\Gamma}(\alpha)}{16(1-\alpha)}\,
  \bigg(\f{E_M}{E_J}\bigg)^2\,\bigg]^{2\alpha/3} 
\end{equation}

\noindent with $\tilde{\Gamma}(\alpha)={\Gamma(\frac{1}{4\alpha})\,\Gamma(\frac{1}{2}+\frac{1}{\alpha})}/[\Gamma(\frac{1}{\alpha})\,\Gamma(\frac{1}{2}+\frac{1}{4\alpha})]$ and 
within the temperature range of validity of Eq.~\eqref{low}. For $T< T_m$ the resistance is an increasing function of temperature whereas it decreases for $T>T_m$. We note that the temperature correction due to $E_M$ is  diverging at zero temperature such that there is a temperature, much lower than $T_m$, below which the perturbative expansion~\eqref{low} is insufficient.

\begin{figure}
\begin{center}
\includegraphics[width=8cm]{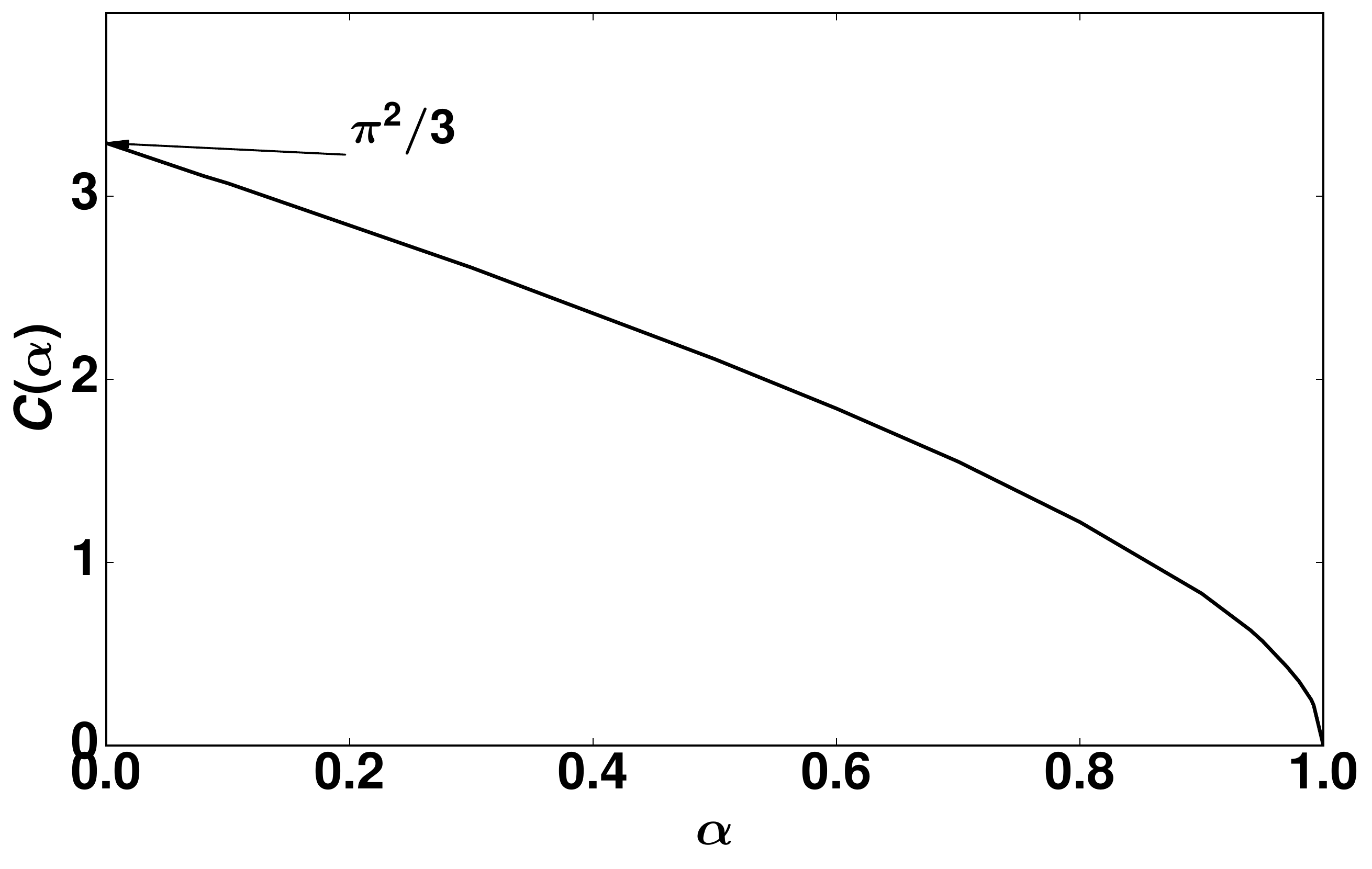}
\caption{$C(\alpha)$ versus the dimensionless dissipation term $\alpha$. The function evolves between $\pi^2/3$ as $\alpha \to 0$ to zero when $\alpha\to 1$. This is in agreement with the result of Fisher and Zwerger~\cite{fisher_zwerger}.
\label{C_alpha}}
\end{center}
\end{figure}

For $T \sim E_C$,  Eq.~\eqref{low} is no longer valid; however we can set $E_M=0$ in Eq.~\eqref{perturbation} since we assume $E_M \ll E_J$. The resulting expression for the resistance exhibits a (local) minimum for
\begin{equation}
  T^* = C(\alpha) \f{2E_C}{\pi^2} \gg T_m 
\end{equation}

\noindent where $C(\alpha)$ can be evaluated numerically and is represented 
in Fig.~\ref{C_alpha}

\begin{figure}
\begin{center}
\includegraphics[width=8cm]{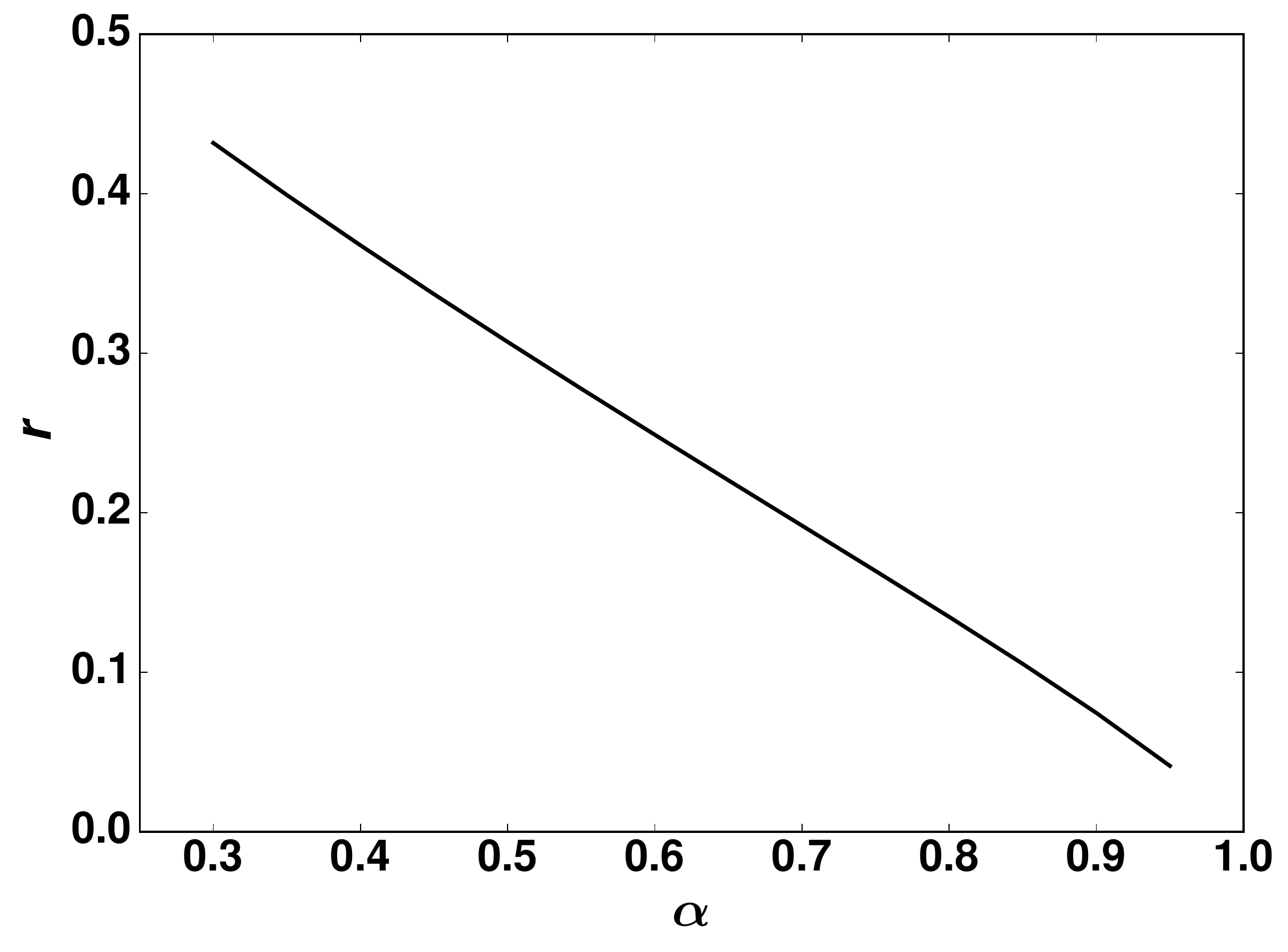}
\caption{Critical ratio $r=(E_M/E_J)_C$, as function of $\alpha$, at which the local maximum $T_m$ and minimum $T^*$ merge. Above, the resistance is a monotonous function of temperature, see also Fig.~\ref{interpolation}. The curve is close to be linear and well fitted by $(E_M/E_J)_C = 0.605-0.590\,\alpha$.
\label{fig_courbe_num}}
\end{center}
\end{figure}

The distance between the local maximum $T_m$ and the local minimum $T^*$ decreases with the ratio $E_M/E_J$. Quite generally for arbitrary $E_M/E_J$, the resistance can be obtained by a numerical evaluation of the integrals in Eq.~\eqref{perturbation}. We thus observe a critical value of $E_M/E_J$, shown in Fig.~\ref{fig_courbe_num} as function of $\alpha$, at which the two extrema meet and disappear. Above this critical value, the resistance becomes a monotonic increasing function of the temperature.

\subsection{Non-perturbative resummation}

We mentioned in the preceding section that perturbation theory fails at low temperature since $E_M$ multiplies a relevant operator. The description of the crossover to very low temperatures thus requires a resummation of the whole perturbation series, and such an exact resummation is not available for general $\alpha$ when both $E_J$ and $E_M$ are non-zero.For $\alpha=1/2$ however, a refermionization technique~\cite{kane_fisher, mora, furusaki} has been successfully applied to compute the crossover for the resistance when $E_J=0$. We extend it below to non-zero $E_J$ where an exact crossover can also be formulated.

The main idea is to interpret 
$\ex{\ir e \hat{\phi}}$ as a bosonized form of a fermion operator
$\hat{\psi}$.
At zero temperature, with a chiral Hamiltonian

\begin{equation}\label{chiral}
H_0=-\ir \displaystyle\int\hat{\psi}^\dagger(x)\,\partial_x\,\hat{\psi}(x)
\end{equation}

\marge the correlation function is 

\begin{equation}\label{correlatorfermion}
\moyenne{\hat{\psi}^\dagger(x,t)\hat{\psi}(0,0)}=
\f{1}{2\pi\ir(t-x)}
\end{equation}

\marge At zero temperature and for $t\gg \tau_s=R\e{s}C$, 
the integral \eqref{correlator} gives

\begin{multline}\label{correlator_RF}
\moyenne{\ex{\ir e\hat{\phi}(t,0)}\,
\ex{-\ir e\hat{\phi}(0,0)}}\simeq \ex{-\ir\pi/(4\alpha)}\times\paf{\tau_s}{\ex{\gamma}\,t}^{1/(2\alpha)}
\end{multline}

\marge For $\alpha=1/2$, the correlator of $\ex{\ir e\hat{\phi}}$ has
the same time dependence as \eqref{correlatorfermion}. The bosonization formula compatible with \eqref{correlator_RF} and 
\eqref{correlatorfermion}
is

\begin{equation}\label{referm}
\hat{\psi}(0)= \sqrt{\f{\ex{\gamma}}{\pi \tau\e{s}}}\,
\hat{a}\,\ex{\ir e \hat{\phi}(0)}
\end{equation}

\marge $\hat{a}$ is a local Majorana fermion with $\hat{a}=\hat{a}^\dagger$
and $\hat{a}^2=1/2$. $\hat{a}$ ensures the anticommutation rules
for the fermionic field $\hat{\psi}(x)$. With the representation 
\eqref{referm}, the quadratic part of Eq.\eqref{hamilto4} can be remplaced
by the Hamiltonian \eqref{chiral}, and

\begin{multline}\label{rM}
-E_M\cos\Big(e\hat{\phi}(0)\Big)=-r_M\,\hat{a}\,\Big(\hat{\psi}(0)-\hat{\psi}^\dagger(0)\Big)\\
\text{with}\quad r_M=E_M\sqrt{\f{\pi \tau\e{s}}{\ex{\gamma}}}
\end{multline}
A straightforward point-splitting calculation connects $\ex{2\ir e\hat{\phi}(0)}$ to $\hat{\psi}(0)\,\partial_x\hat{\psi}(0)$, since $\hat{\psi}(0) \hat{\psi}(0)=0$ due to Fermi statistics. The two operators have scaling dimension $2$ when $\alpha=1/2$. The precise connection is obtained by identifying the two-point correlators using Eq.\eqref{correlatorfermion}, with the result
\begin{multline}
-E_J\cos(2e\hat{\phi}(0))=\ir r_J\,\Big(\hat{\psi}(0)\,\partial_x\hat{\psi}(0)
+\hat{\psi}^\dagger(0)\,\partial_x\hat{\psi}^\dagger(0)\Big)\\
\text{with} \quad r_J=\f{\pi\,{\tau\e{s}}^2\, E_J}{\ex{2\gamma}}
\end{multline}
The refermionized Hamiltonian takes the form
\begin{multline}\label{hamiltoRF}
H=-\ir \displaystyle\int\hat{\psi}^\dagger(x)\,\partial_x\,\hat{\psi}(x)
-r_M\,\hat{a}\,\Big(\hat{\psi}(0)-\hat{\psi}^\dagger(0)\Big)\\
+\ir r_J\,\Big(\hat{\psi}(0)\,\partial_x\hat{\psi}(0)
+\hat{\psi}^\dagger(0)\,\partial_x\hat{\psi}^\dagger(0)\Big),
\end{multline}
which is quadratic and exactly solvable. This effective Hamiltonian allows for a complete
resummation for energies smaller than $E_C$.
Interestingly, Eq.\eqref{hamiltoRF} already appeared in a different context~\cite{lin, fidkowski, zuo, pikulin, sela} as it can represent a semi-infinite helical wire - unfolded as a chiral mode on an infinite line - coupled at $x=0$  to a topological superconductor hosting a single Majorana bound state at its edge. $r_M$ plays the role of the tunnel coupling to the Majorana bound state while $r_J$ generates Andreev reflections at  the superconductor. In this model, an incoming electron can be reflected as an electron or 
an hole (and vice versa).

Eq.\eqref{hamiltoRF} is easily diagonalized using a mode expansion~\cite{fidkowski, sela} for $\hat{\psi}(x)$ and $\hat{\psi}^\dagger(x)$ summarized in Appendix~\ref{appen-smatrix}. Moreover, we can describe the relation between the left-moving electrons/holes $(x<0)$ and the right-moving electrons/holes $(x>0)$ with the S-matrix. At the formal level, the second term in the expression~\eqref{resistance} of the resistance, involving the correlator $\langle\hat{f}(\tau)\hat{f}(0)\rangle$, coincides with the differential conductance of the boundary helical model~\cite{fidkowski}. Hence, the resistance~\eqref{resistance} can be expressed using one of the S-matrix components:
\begin{equation}\label{resistmatrix}
\f{R}{R\e{s}}=1-\displaystyle\int_{-\infty}^{+\infty}\dd\omega\,\bigg(-
\f{\partial n_f}{\partial \omega}\bigg)\,|S\e{ph}(\omega)|^2
\end{equation}
where $n_f(\omega) = (1+e^{\beta \omega})^{-1}$ is the Fermi distribution and 
$S\e{ph}$ is the probability for an incoming electron with energy $\omega$
to be reflected as a hole.  The derivation
of $S\e{ph}$ is reproduced in Appendix~\ref{appen-smatrix}:
\begin{equation}\label{smatrix}
S\e{ph}(\omega)=
\f{\ir (2r_J\,\omega^2+{r_M}^2)}{\ir{r_M}^2+\omega\,(1+r_J{r_M}^2+{r_J}^2\omega^2)}
\end{equation}
in agreement with Ref.~\onlinecite{fidkowski}. Inserting Eq.\eqref{smatrix} into Eq. \eqref{resistmatrix}, 
the effective resistance reads

\begin{equation}\label{resistdimension}
\f{R}{R\e{s}}=1-F\bigg(\f{4\ex{\gamma}}{\pi^2}\,\f{TE_C}{{E_M}^2},
\f{\pi^5}{8\ex{3\gamma}}\,\f{{E_M}^2E_J}{{E_C}^3}\bigg)
\end{equation}

\marge where the dimensionless function $F(\tilde{T},\tilde{r})$
is given by

\begin{equation}\label{dimensionless}
F(\tilde{T},\tilde{r})=\displaystyle\int_{-\infty}^{+\infty}
\f{\dd x}{2\cosh^2x}\,\f{(2\tilde{T}^2\,\tilde{r}\,x^2+1)^2}{1+\tilde{T}^2x^2
\,(1+\tilde{r}+\tilde{T}^2\,\tilde{r}^2x^2)^2}
\end{equation}


\marge For $E_J=0$ , the integration can be performed and we recover the known result~\cite{shon_zaikin, weiss_wollensak}

\begin{equation}
\f{R}{R\e{s}}\bigg|_{E_J=0}=1-\f{\pi {E_M}^2}{4\ex{\gamma} E_C\,T}\,
\Psi'\bigg(\f{1}{2}+\f{\pi {E_M}^2}{4\ex{\gamma} E_C\,T}\bigg)
\end{equation}



\noindent At zero temperature, one obtains $F=1$ consistent with a fully coherent Josephson junction. Let us emphasize that the result~\eqref{resistdimension} was obtained  assuming $T\ll 2E_C/\pi$ and $E_J,E_M\ll E_C$.
 As a consequence, we have
${E_M}^2E_J/{E_C}^3\ll 1$ such that second parameter $\tilde{r}$ in Eq.\eqref{dimensionless} is always much smaller than one. At low temperature $T\ll {E_M}^2/E_C$, we can therefore ignore $E_J$ to obtain the asymptotic expression
\begin{equation}\label{asymptotic}
  \f{R(T)}{R\e{s}}
  \simeq \f{4\ex{2\gamma}}{3\pi^2}\,
\paf{T}{{E_M}^2/E_C}^2.
\end{equation}
In the opposite limit  $T\gg {E_M}^2/E_C$, we keep $\tilde{r} \sim E_J$ and recover the result Eq.~\eqref{low} of the preceding section for $\alpha=1/2$.

\begin{figure}
\begin{center}
\includegraphics[width=8cm]{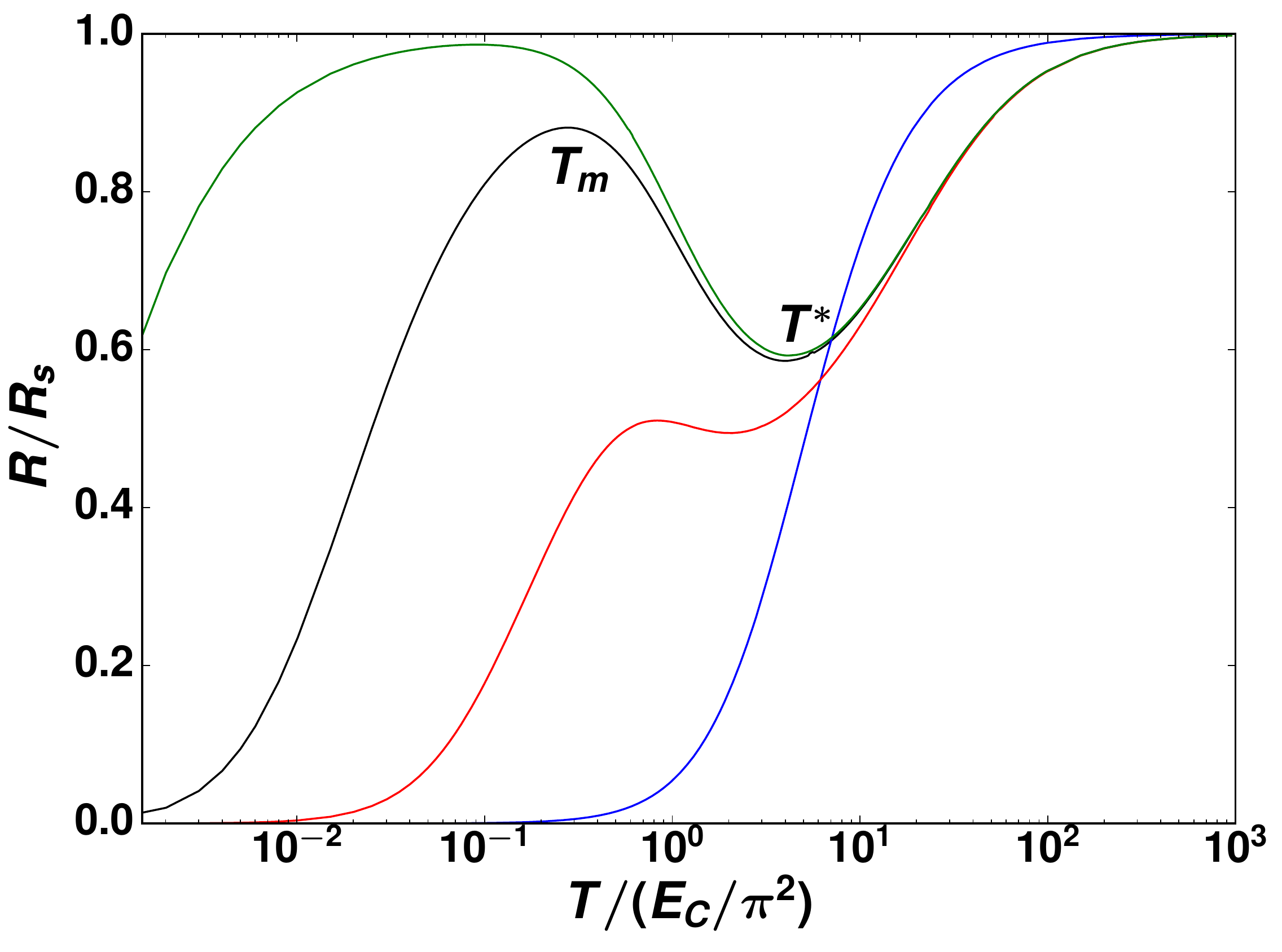}
\caption{Dimensionless resistance $R/R\e{s}$ as function of the reduced temperature $T/(E_C/\pi^2)$ for $\alpha=1/2$ and different values of  $(E_J/E_C,E_M/E_C)$ obtained by interpolation between Eq.~\eqref{resistdimension} and Eq.~\eqref{perturbation}. We represent $(E_J/E_C,E_M/E_C) = (0.8,0.1)$ (red line), $(0.03,0.5)$ (blue line), $(0.8, 0.006)$ (green line), $(0.8,0.03)$ (black line). The two extrema $T_m$ and $T^*$ merge at $E_J/E_M = 0.31$, see Fig.~\ref{fig_courbe_num}.
\label{interpolation}}
\end{center}
\end{figure}

We plot in Fig.~\ref{interpolation} the interpolation between  Eq.~\eqref{resistdimension} and Eq.~\eqref{perturbation} for different values of ($E_M/E_C, E_J/E_C)$. We thus observe the crossover where the two extrema disappear.

\section{Large Josephson energy}


The analysis of Sec.~\ref{sec-coulomb} was restricted to the Coulomb blockade regime where the bare charging energy $E_C$ is the largest energy scale. Coulomb blockade tends to pin the superconductor charge which has the effect of delocalizing the conjugated phase variable. The shunted Josephson junction is then closer to an insulator, with a differential resistance below but in the vicinity of $R_s$, except at very low temperature where the relevant Josephson energy $E_M$ takes over and reestablishes a dissipationless Josephson tunneling. 

The resulting resistance, shown in Fig.~\ref{interpolation}, exhibits a local minimum at $T = T^* \sim E_C$ with a distance to the fully shunted junction $R=R_s$ increasing with $E_J$, reflecting a partial relocalization of the phase. This scaling suggests that the local minimum keeps decreasing with $E_J/E_C$ until it reaches an almost vanishing resistance in a certain temperature range for $E_J$ larger than $E_C$. In what follows, we consider directly the regime of deep potential wells $E_J \gg E_C$ while $E_M$ is chosen below the plasma frequency $\omega\e{p}=\sqrt{8E_JE_C}$.

We first diagonalize the model in the absence of the ohmic environment in Sec.~\ref{sec-dissipationless} and then take in Sec.~\ref{sec-dissipation} $\alpha=1/2$ where a mapping to the Emery-Kivelson model can be demonstrated. This gives the exact resistance for $\alpha=1/2$ and a qualitative picture for $\alpha$ between $1/4$ and $1$, extending the analysis of Sec.~\ref{sec-coulomb}.

\subsection{Dissipationless case}\label{sec-dissipationless}

In the absence of dissipation, the Hamiltonian~\eqref{hamilto} simplifies as $H = H_0 - E_M \cos (\hat{\varphi}/2)$
with the transmon Hamiltonian~\cite{koch}
\begin{equation}\label{hamiltonian-dissipationless}
  H_0(n_g) =4E_C\,(\hat{N}-n_g)^2-E_J \cos\hat{\varphi}
\end{equation}
where $n_g$  is the offset charge of the capacitor. In the phase representation, $\hat{N}  =i \partial_\varphi$ is acting on $2 \pi$-periodic functions, and $H_0$ can be diagonalized exactly using Mathieu functions~\cite{wilkinson}. For $E_J \gg E_C$, the energy of the ground state takes the suggestive form
\begin{equation}\label{dispersion}
E_0 (n_g) = \f{\omega\e{p}}{2} - t_0 \cos \left(2 \pi n_g \right),
\end{equation}
where~\cite{koch}
\begin{equation}
  t_0 = 16\,\sqrt{\frac{E_JE_C}{\pi}}\,
\left(\frac{E_J}{2E_C}\right)^{1/4}\exp\left(-\sqrt{\frac{8E_J}{E_C}}\right)
\end{equation}
The ground state energy has a periodicity of one in $n_g$ as expected from the discreteness of $\hat N$.

Although the above derivation is self-contained, it is instructive to formulate it using a Bloch band description~\cite{shon_zaikin, catelani}. The Hamiltonian $H_0$~\eqref{hamiltonian-dissipationless} with the compact phase $\hat \varphi$ in $[0,2\pi]$ is mathematically equivalent to solving $H_0(0)$ with an extended phase, {\it i.e.}  $\hat \varphi$ between $-\infty$ and $\infty$, with $n_g$ playing the role of the quasimomentum. In this langage, Eq.~\eqref{dispersion} as function of $n_g$ is a band dispersion. Focussing again on the regime $E_J\gg E_C$, the wavefunctions of low-energy eigenstates are strongly localized near the minima of the cosine potential and the Wannier function of the lowest band is the ground state of an harmonic oscillator,
\begin{equation}
  W_0(\varphi)=\left(\frac{E_J}{8\pi^2 E_C}\right)^{1/8}\,\exp\left(-\frac{\varphi^2}{2}\sqrt{\frac{E_J}{8E_C}}\right),
\end{equation}
 with energy $\omega\e{p}$. The small overlap between consecutive Wannier functions induces a nearest-neighbor hopping term $t_0/2$. We thus obtain a tight-binding model whose diagonalization reproduces Eq.~\eqref{dispersion} and $t_0$ is identified as the bandwidth of the lowest band in the cosine potential.

\begin{figure}
\begin{center}
\includegraphics[width=6cm]{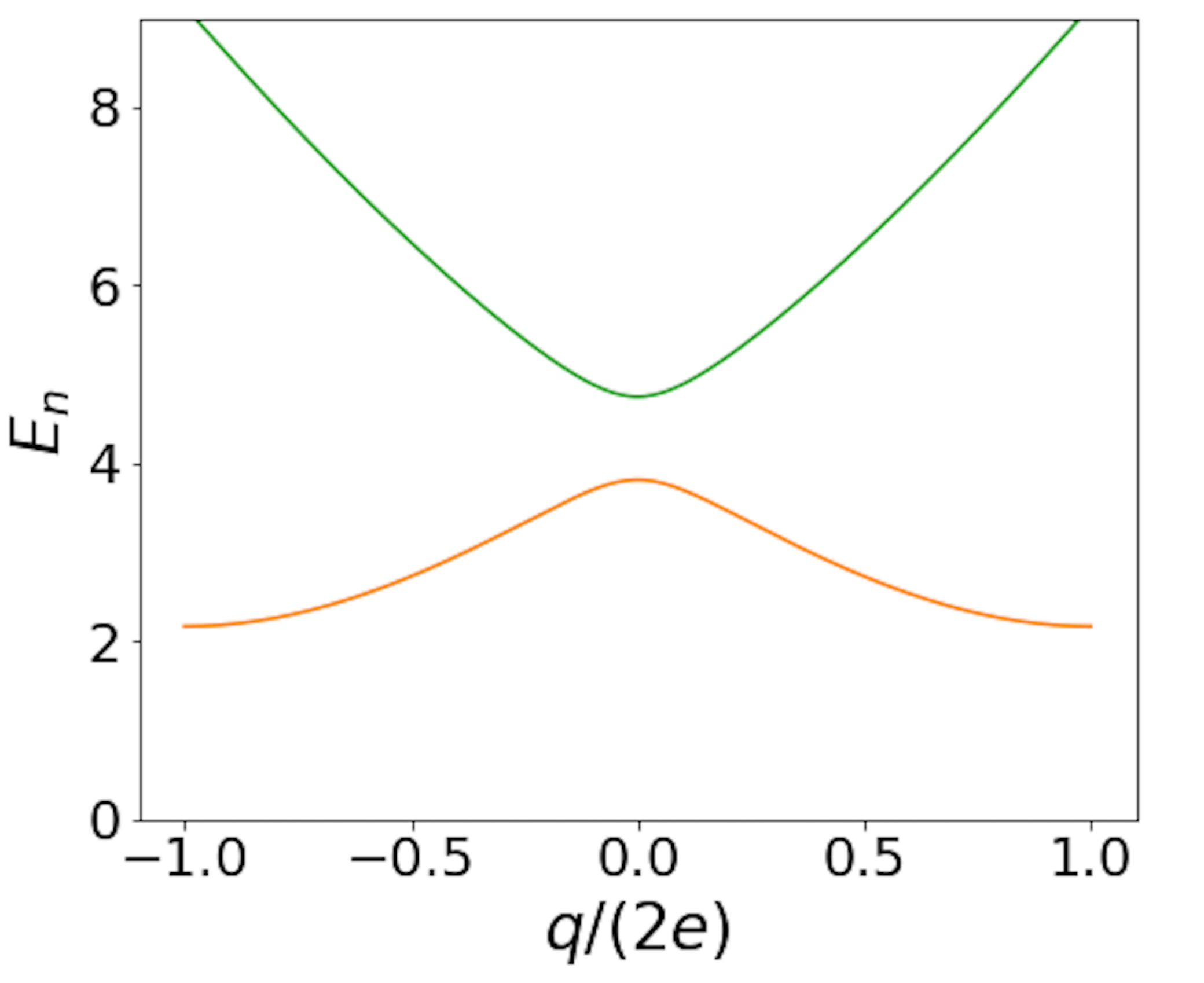}
\includegraphics[width=6.88cm]{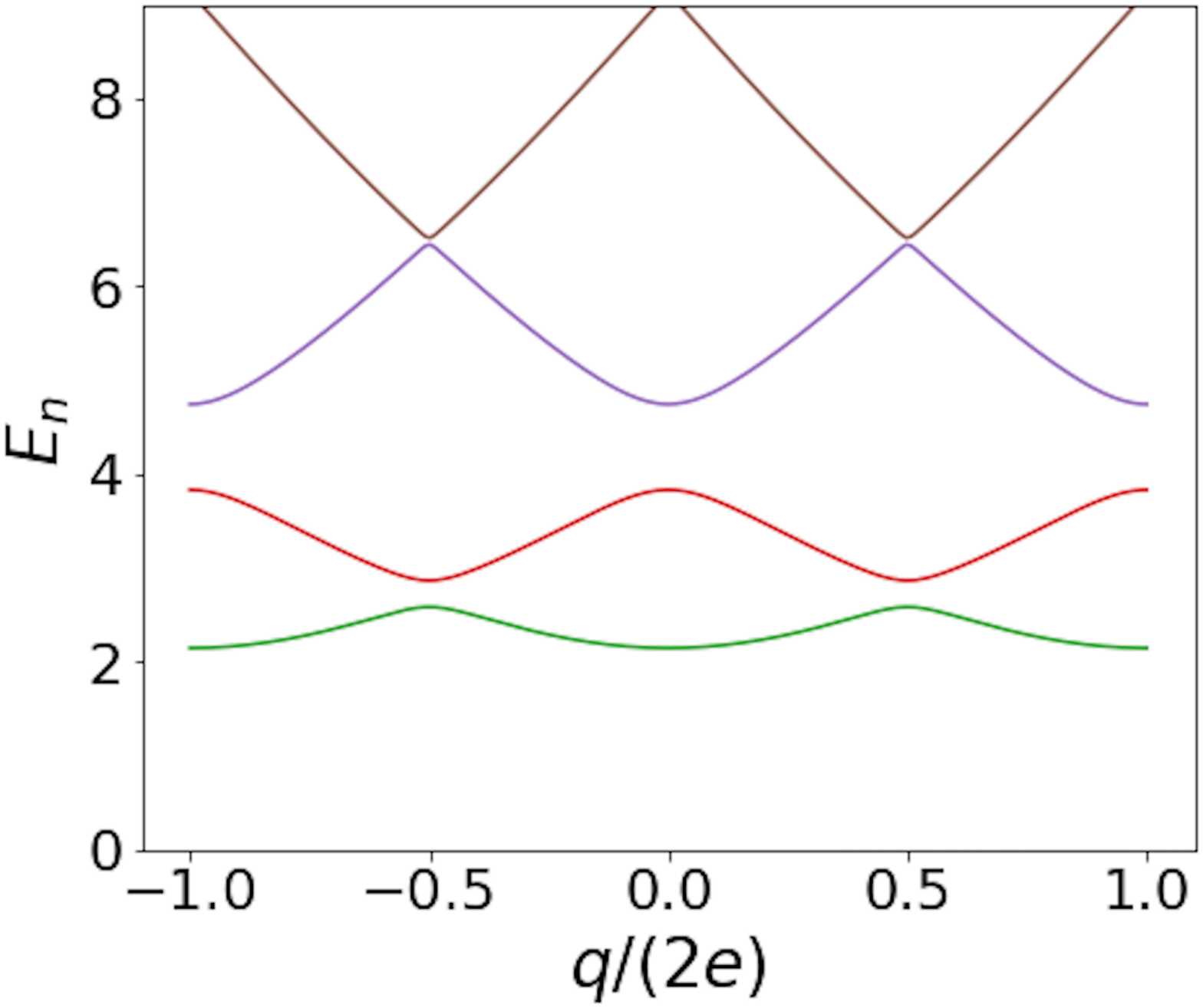}
\caption{Lowest bands obtained by a numerical solution of $H$ as a function of the charge offset $n_g$ in the absence of dissipation. The numerical solution is found~\cite{qutip1, qutip2} 
by performing a truncation in the charge basis $|n\rangle$ (eigenstates of $\hat{N}$) and diagonalizing a finite-size version of $H$. In both pictures, $E_C=1$ and $E_J=3$.
(a) Spectrum for $E_M=0$. (b) Spectrum for $E_M=3$. Gaps open at the edges of the reduced Brillouin zone.
 \label{fig_TB_num}}
\end{center}
\end{figure}

 Next, we include $E_M$ and numerically evaluate the spectrum of $H$. The presence of $E_M$ doubles the size of the unit cell folding the spectrum at $n_g = \pm 1/2$, corresponding to single-electron tunneling, and opening gaps at the edge of the new Brillouin zone as illustrated in Fig.\ref{fig_TB_num}. In order to make further analytical progress, we consider $E_M < \omega\e{p}$ such that the different bands of $H_0$ are not mixed, and project the Hamiltonian onto the lowest band. Due to non-zero $E_M$, the wavefunctions must have a periodicity of $4 \pi$ in $\varphi$. For a given charge offset $n_g$, we find two such functions in the lowest band~\cite{yavilberg, ginossar}:
\begin{equation}\label{basis}
  \Psi_{\pm,n_g}  (\varphi)= \sum_{p \in \mathbb{Z}} (\pm 1)^p \, W_0(\varphi- 2 p \pi) e^{i n_g (2 p \pi - \varphi)} 
\end{equation}
with energies $E_0(n_g)$ and $E_0 (n_g+1/2)$. These are $4\pi$-periodic functions, even and odd with respect to the parity operator $(-1)^{\hat N} \Psi_{\pm,n_g} = \pm \Psi_{\pm,n_g}$. After projecting the Hamiltonian $H$ onto the basis~\eqref{basis}, we get
\begin{equation}\label{low-ener}
  \begin{split}
    H_{LE} (n_g) & = t_0 \cos \left(2 \pi n_g \right) \sigma_z + c_0 E_M \sigma_x \\
    & + d_0 E_M  \sigma_y \sin \left(2 \pi n_g \right),
  \end{split}
\end{equation}

\noindent where $\sigma_i$ ($i=x, y, z$) are Pauli matrices operating in parity space and the constant term $\omega\e{p}/2$ has been removed. This derivation of the different overlaps in Eq.~\eqref{low-ener} uses that $W_0 (\varphi)$ takes appreciable values only close to $\varphi \simeq 0$. One consequence is that $E_M$ does not enter the diagonal elements - or only with a very small contribution neglected here, whereas there is a perfect overlap $c_0=1$ along $\sigma_x$. The overlap along $\sigma_y$ is given by
\begin{equation}
  d_0 = \frac{2^{1/3}t_0}{3^{-1/3}\omega_p}\,\sqrt{\frac{E_C}{E_J}}\,\Gamma\paf{2}{3} \ll 1,
\end{equation}
and can be also neglected. The $\sigma_x$ and $\sigma_y$ components in Eq.~\eqref{low-ener} can be seen respectively as a staggered potential and a staggered hopping amplitude in the tight-binding model. We note that applying a non-zero flux between the two Josephson junctions can change the relative values of $c_0$ and $d_0$. The limiting case $c_0=0$ and $d_0=1$ corresponds to the SSH model~\cite{ssh}.

\subsection{Effective Hamiltonian with dissipation}\label{sec-dissipation}

The projection to the lowest band of the extended potential can still be applied to the original Hamiltonian~\eqref{hamilto} provided the ohmic dissipation is not too strong. It amounts to an adiabatic approximation where one replaces the charge offset by $2 e n_g = - \hat{Q} - \int^{t}\dd t'\,I_0(t')$ in the transmon Hamiltonian $H_0$~\eqref{hamiltonian-dissipationless}. It is justified as long the input current $I_0$ is weak and varies slowly in time, and if $\alpha^2 \ll 2 \pi^2 E_J/E_C$~\cite{korshunov}.

The projected Hamiltonian is
\begin{equation}
 H_P = H_{LE} \left( \f{\hat{Q}}{2 e} + \f{Q_0 (t)}{2 e} \right)  + H_{R\e{s}},
\end{equation}
where $Q_0 (t) = \int^{t}\dd t'\,I_0(t')$, which we expand to first order in $Q_0$ as
\begin{equation}\label{tight_model3}
  H_P \simeq H_{LE} \paf{\hat{Q}}{2 e}   + H_{R\e{s}} - t_0 \frac{\pi Q_0 (t)}{e} \sin\paf{\pi \hat{Q}}{e},
\end{equation}
and, for $Q_0=0$,
\begin{equation}\label{tight_model4}
  H_P = t_0\cos\Big(\f{\pi}{e}\hat{Q}\Big)\,\sigma_z + E_M\,\sigma_x
  +  H_{R\e{s}}
\end{equation}

Due to the absence of the capacitive term, the mode expansion of the fields in the transmission line differs
slightly from Sec.~\ref{sec-unitary} and we have
\begin{subequations}
\begin{align}
  \hat{\phi}(0) & =\sqrt{\f{R\e{s}}{4\pi}}\,\displaystyle\int_{0}^{\omega\e{p}}
\f{\dd\omega}{\sqrt{\omega}}\,\bigg[\f{2}{1-\ir\tau_0\omega}\,
  a\e{in,$\omega$}+ \text{h.c}\bigg], \\[2mm]
  \hat Q & = \frac{1}{\sqrt{4 \pi R\e{s}}} \int_0^{\omega\e{p}} \f{d \omega}{\sqrt{\omega}} \left[ \f{-2 i \tau_0 \omega}{1-\ir\tau_0\omega} a_{\rm in,\omega}  - {\rm h.c.} \right],
\end{align}
\end{subequations}
where $\tau_0$ is a regularizing time that is eventually sent to infinity. Frequencies are cut off at the plasma frequency $\omega\e{p}$ at which higher bands start to play a role. The correlator $K_{\hat Q} (t) = \langle \ex{i \pi\hat{Q}(t)/e}\ex{-i \pi\hat{Q}(0)/e}\rangle$ describes now charge fluctuations. At zero temperature, one gets
\begin{equation}
  K_{\hat Q} (t) = \exp \left[ 2 \alpha \int_0^{\omega\e{p}} \f{d \omega}{\omega} \f{(\tau_0 \omega)^2}{1+(\tau_0 \omega)^2} \left( e^{- i \omega t} -1 \right)  \right],
\end{equation}
or, for $1/\omega\e{p} \ll t \ll \tau_0$ and $\alpha$, $K_{\hat Q} (t) \sim 1/(\omega\e{p} t)$. The same fermionization as Eq.~\eqref{referm} can be performed such that
\begin{equation}
t_0\,\cos\Big(\f{\pi}{e}\,\hat{Q}\Big)=r_0\,\hat{a}\,
 \Big(\hat{\psi}(0)-\hat{\psi}^\dagger(0)\Big)
\end{equation}
with $r_0 \sim t_0/\sqrt{\omega\e{p}}$. The Hamiltonian~\eqref{tight_model4} can be further transformed using the representation of Pauli matrices in terms of Majorana fermions~\cite{pauli}, \textit{i.e} $\sigma_x=\ir \eta_2\eta_3$, $\sigma_y=\ir\eta_3\eta_1$ and
$\sigma_z=\ir\eta_1\eta_2$. The commutation relations of the Pauli matrices are ensured by the Clifford algebra $\{\eta _i,\eta_j\}=2\delta_{ij}$. Using these representations, the Hamiltonian~\eqref{tight_model4} becomes
 \begin{multline}\label{refermTB}
 H_P=-\ir\displaystyle\int\psi^\dagger(x)\partial_x\psi +\ir E_M\,\eta_2\eta_3 \\-\ir r_0\,
 \Big(\hat{\psi}(0)-\hat{\psi}^\dagger(0)\Big)\,\hat{a}\,\eta_1\eta_2.
 \end{multline}
 $\hat{a}\, \eta_1$ commutes with $H_P$ such that we can choose $\hat{a} \, \eta_1=\ir/\sqrt{2}$ and,
 \begin{multline}\label{refermTB2}
 H_P=-\ir\displaystyle\int\psi^\dagger(x)\partial_x\psi +\ir E_M\,\eta_2\eta_3 \\ + \f{r_0}{\sqrt{2}}
 \Big(\hat{\psi}(0)-\hat{\psi}^\dagger(0)\Big)\, \eta_2.
 \end{multline}
This Hamiltonian coincides exactly with the effective model found by Emery and Kivelson~\cite{emer-kivel,georges} to solve the two-channel Kondo model in the presence of a magnetic field. $r_0$ is a relevant operator driving the system to a strong-coupling fixed point where $r_0$ is large - or $E_J$  large - and the Majorana fermion $\eta_2$ is screened. The effect of the magnetic field $\sim E_M$ is to stop the renormalization group flow. Eq.~\eqref{refermTB2}  is quadratic and therefore analytically solvable.
 
 \subsection{Resistance}

 The voltage drop across the junction is 
 \begin{equation}\label{TB_voltage}
   \begin{split}
   V& =\langle \dot{\hat{\phi}}(t)\rangle=\ir \langle[H_P,\hat{\phi}]\rangle \\[2mm]
   & = -t_0  \f{\pi}{e} \langle\sin\Big(\f{\pi}{e}\hat{Q}\Big)\,\sigma_z \rangle + \ir \langle[H_{R\e{s}},\hat{\phi}]\rangle
   \end{split}
 \end{equation}
 where we use again the notation $\hat{\phi}(t)=\hat{\phi}(x=0,t)$. The second term $[H_{R\e{s}},\hat{\phi}]$  gives zero. Using the Kubo formula and Eq.~\eqref{tight_model3} for the linear coupling to the bias current, we obtain the expression for the resistance
\begin{equation}\label{TB_resistance}
 \f{R(T)}{R\e{s}} = 2\pi\,\alpha \, \limt_{\omega\to 0}
 \,\mathcal{R}\Big(\f{\ir}{\omega}\,\limt_{\ir\omega_n\to \omega+\ir 0^+}
 \,\displaystyle\int_{0}^{\beta}\,\ex{\ir\omega_n\tau}\langle\hat{g}(\tau)\hat{g}(0)\rangle\Big)
 \end{equation}
 where
 \begin{multline}\label{TB_resistance2}
\hat{g}(\tau)=\ir t_0\,\sin\Big(\f{\pi}{e}\,\hat{Q}(\tau)\Big)\,\eta_1(\tau)\,\eta_2(\tau)\\[2mm]
= -\ir\f{r_0}{\sqrt{2}}\,\Big(\hat{\psi}(\tau)+\hat{\psi}^\dagger(\tau)\Big)\,\eta_2(\tau)
 \end{multline}
 where $\hat{\psi}(\tau)=\hat{\psi}(\tau,x=0)$.
 Using the S-matrix formalism introduced in the previous section with Eq.\eqref{refermTB}, we finally obtain
 \begin{equation}\label{mobility2}
\f{R(T)}{R\e{s}} =\displaystyle\int_{-\infty}^{+\infty}\dd\omega\,\bigg(-\f{\partial n_f}{\partial \omega}\bigg)
 \f{{r_0}^4\,\omega^2}{(\omega^2-4{E_M}^2)^2+{r_0}^4\,\omega^2}
 \end{equation}
 For $E_M = 0$, one obtains~\cite{weiss_wollensak}

\begin{equation}\label{TB_EM_0}
\f{R}{R\e{s}}=\f{{r_0}^2}{2\pi T}\,\Psi'\Big(\f{1}{2}+\f{{r_0}^2}{2\pi T}\Big)
\end{equation}

\noindent with the infrared insulating fixed point, $R=R_s$, at zero temperature.

\begin{figure}
\begin{center}
\includegraphics[width =0.9\columnwidth]{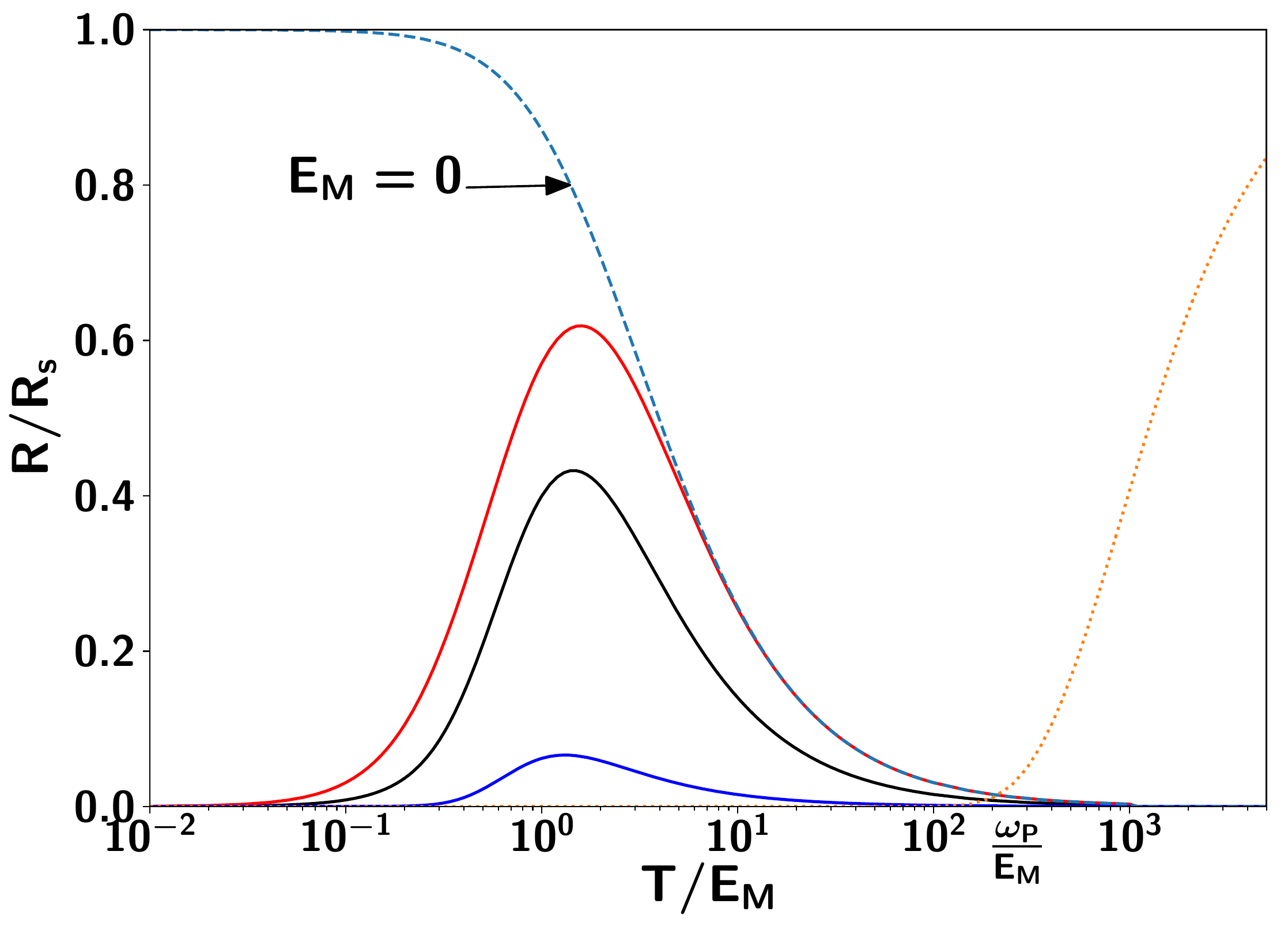}
\caption{Resistance versus temperature for ${r_0}^2/ 2 E_M= 0.1$ (blue), $1$ (black), $2$ (red), where ${r_0}^2/ E_M \sim {t_0}^2/ \omega_p E_M$. The limiting case of vanishing $E_M$ is shown in dotted line for comparison. The analysis in this work is limited to temperatures lower than the plasma frequency $\omega_p$. Above, a thermally actived behaviour $\sim e^{-E_J/T}$ towards a full resistance $R=R_s$ occurs~\cite{nozieres} (dotted yellow line) corresponding to a complete delocalization of the phase.
\label{fig_resistanceTB}}
\end{center}
\end{figure}

A non-zero $E_M$ drastically changes to the infrared superconducting fixed point. Eq.~\eqref{mobility2} gives the resistance $R=0$ at zero temperature. The temperature dependence of the resistance is shown in Fig.\ref{fig_resistanceTB} by evaluating the integral in Eq.~\eqref{mobility2}.
For $T \gg E_M$, the integral simplifies as
\begin{equation}\label{TB_limit}
\f{R}{R\e{s}}\simeq \f{\pi}{4}\,\f{{r_0}^2}{T}
\end{equation}

We obtain the same expression as Eq.~\eqref{TB_EM_0} in the high-temperature limit:
for $T\gg E_M$, the resistance is only controlled by ${r_0}^2$.

The approach discussed in this section is limited to temperatures below the plasma frequency $\omega\e{p}$. Above $\omega\e{p}$, the phase delocalizes via  thermal activation across the minima of the deep Josephson potential~\cite{nozieres}, and the resistance increases again until it reaches the insulating regime $R=R_s$ for temperature much larger than $E_J$.
  

 \section{Conclusion}

 In this paper, we have considered a superconducting junction with a Josephson energy being the sum of two contributions: the usual $2 \pi$-periodic energy and an additional $4 \pi$-periodic energy which can be the sign of a topological junction hosting Majorana bound states. The model accounts for a single junction with the two periodicities or, alternatively, two separated Josephson junctions in parallel. The existence of the two commensurate phase periodicities is not necessarily a sign of topology but could also occur with, for instance, a Kondo impurity, a Josephson rhombus device or in hybrid ferromagnetic-superconducing junctions.

 A sufficiently strong shunt resistor in parallel with the junction drives the device via a quantum phase transition to an insulating regime where the zero-temperature Josephson current is suppressed. We have focussed on values of the resistance such that the $2 \pi$-periodic term alone would give an insulating state while the $4 \pi$-periodic energy is still superconducting, resulting in a competition between the two terms. We have derived the corresponding non-monotonic behaviour of the total differential resistance of the device as a function of temperature.

 For a charging energy much larger than the Josephson energy, the junction is essentially resistive at high temperature. The resistance increases with temperature for $T > E_C$ but decreases for  $T < E_C$, reaching a minimum at $T = T^* \sim E_C$. For even lower temperatures, a crossover towards a fully superconducting junction with zero resistance was established below a local temperature maximum $T_m$. These features were first derived using perturbation theory and then an exact analytically expression was obtained when $\alpha=1/2$, or $R_s=2R\e{q}$. The exact refermionization maps the model onto an helical one-dimensional wire coupled to a topological superconductor.

 When the charging energy is the smallest energy scale, we confirmed the non-monotonic behaviour with a tight-binding approach connecting the Josephson wells. Delocalizing the superconducting phase, the hopping between the wells is suppressed by dissipation - by duality, the coupling of the phase to dissipation decreases with the shunt resistance - and a superconducting behaviour is restored  at low energy. 
 For $\alpha=1/2$, or $R_s=2R\e{q}$, we find a mapping to the Emery-Kivelson line of the two-channel Kondo model under finite magnetic field. It provides again an exact expression for the temperature-dependent differential resistance.
 
A straightforward extension of our work is the study of the non-linear current-voltage characteristic where the weight of the Josephson peak is transferred by the environment to higher voltages~\cite{didier}. Also our analysis is restricted to an equilibrium electromagnetic environment~\cite{torre2012} and the prospect of exciting photons around the Josephson junction~\cite{mendes2015} offers an appealing direction of research.

\acknowledgments

This work has benefited from fruitful discussions with P. Joyez and A. Murani.


\begin{appendix}


\section{Linear response theory}\label{appen-linear}
 
We set $E_M=0$ to simplify the discussion but without loss of generality.
 The contribution $-I_0 (t)\,\hat{\phi}(0)$ in the Hamiltonian~\eqref{hamilto4} can be regarded as a perturbation. We  use the framework of linear response theory to compute the voltage drop across the junction. The Kubo formula gives
 
\begin{multline}\label{appendix_voltage}
V(t)=\moyenne{\dot{\hat{\phi}}}=\ir\,\displaystyle\int_{-\infty}^{t}\,\dd t'\,
 \moyenne{\big[\dot{\hat{\phi}}(t),\hat{\phi}(t')\big]}\,I_0(t')\\
 =-\displaystyle\int_{-\infty}^{+\infty}
 \dd t'\,G_{\dot{\phi},\phi}^\text{ret}(t-t')\,I_0(t')
 \end{multline}

\noindent where $\hat{\phi}(t)=\hat{\phi}(x=0,t)$ and the retarded Green's function 
\begin{equation}
G^\text{ret}(t-t')=-\ir\theta(t)
 \moyenne{\big[\dot{\hat{\phi}}(t),\hat{\phi}(t')\big]}
 \end{equation}

 \marge where $\theta(t)$ is the Heaviside function. In Fourier space, Eq.\eqref{appendix_voltage} becomes $V(\omega)=-G^\text{ret}_{\dot{{\phi}},{\phi}}(\omega)\,I_0(\omega)$.
 $G^\text{ret}(\omega)$ is analytical in the upper complexe semiplane only. As a consequence, for a real argument $\omega$, the limit $\omega+\ir 0^+$ has 
to be considered. Now, we use the notation $\phi_0=2e\,\hat{\phi}(0)$.
With 
$G_{\dot{\phi},\phi}^\text{ret}(\omega)=-\ir\omega\,G_{{\phi},\phi}^\text{ret}(\omega)$, the linear resistance is 

\begin{equation}\label{appendix_resistance}
R(\omega)=R\e{s}\,\f{\alpha}{2\pi}\times \ir\omega\,G_{{\phi_0},\phi_0}^\text{ret}(\omega+\ir0^+)\end{equation}

\noindent In the last expression, we have used 
$1/4e^2=R\e{s}\,\alpha/2\pi$. To compute $G^\text{ret}(\omega+\ir 0^+)$, we compute 
another correlation function in the imaginary time formalism and do an 
analytic continuation. We can compute the Matsubara Green's function 
$G(\tau)=-\moyenne{\mathcal{T}_\tau {\phi_0}(\tau){\phi}_0(0)}$ where 
$\mathcal{T}_\tau$ is the time-ordering. In this article, we want to compute the DC resistance ($\omega\to 0$) so we need to take the real part of Eq.\eqref{appendix_resistance}. The linear resistance becomes

\begin{multline}\label{appendix_resistance1}
\f{R(T)}{R\e{s}}=-\f{\alpha}{2\pi}\, \limt_{\omega\to 0}\mathcal{R}\bigg[\ir\omega\times\\
\limt_{\ir\omega_n\to \omega+\ir 0^+}\,
\displaystyle\int_{0}^{\beta}\dd\tau\,\ex{\ir\omega_n\tau}\,\moyenne{{\phi_0}(\tau){\phi}_0(0)}\bigg]
\end{multline}

\noindent In the last expression the time-ordering factor doesn't appear because the $\tau$ is between $0$ and $\beta>0$. We can use an other expression of \eqref{appendix_resistance1} where the correlator is written with matsubara frequencies:

\begin{equation}\label{appendix_resistance2}
\f{R(T)}{R\e{s}}=-\f{\alpha}{2\pi}\, \limt_{\omega\to 0}\mathcal{R}\bigg[\ir\omega\,
\limt_{\ir\omega_n\to \omega+\ir 0^+}\,
\f{1}{\beta}\,\SUM{\ir\omega_k}{}\,\moyenne{{\phi}_0(\ir\omega_n){\phi}_0(\ir\omega_k)}\bigg]
\end{equation}

\noindent The analytical expression  of the linear resistance is only determined by the correlator $\moyenne{{\phi}_0(\tau){\phi}_0(0)}$. It can be computed with the Euclidian action~\cite{shon_zaikin,kane_fisher,falci,korshunov,schmid}
\begin{multline}\label{appendix_action}
S=\f{1}{2\beta}\,\SUM{\ir\omega_n}{}\Big(\f{1}{8E_C}\,{\omega_n}^2+\f{\alpha}{2\pi}\,|\omega_n|\Big)\,|\phi_0(\ir\omega_n)|^2\\
-E_J\,\displaystyle\int_{0}^{\beta}\dd\tau\,\cos\phi_0(\tau)
\end{multline}
obtained from Eq.~\eqref{hamilto4} as Eq.~\eqref{action} in the main text.

\noindent In order to have a convenient expression of the correlator, we add to 
the action \eqref{appendix_action} a source term of the form~\cite{kane_fisher}  

\begin{equation}\delta S=\f{1}{2\beta}\,\SUM{k}{}\Big(\f{\alpha}{2\pi} |\omega_k|+\f{1}{8E_C}\,{\omega_k}^2\Big)\,a(-\ir\omega_k)\,
\phi_0(\ir\omega_k)\end{equation}

\noindent Our new action becomes $S\e{tot}=S+\delta S$. We introduce the notation $k_\alpha=\alpha/(2\pi)$ and $k_C=1/(8E_C)$. With the action 
$S\e{tot}$, the correlator is

\begin{multline}\label{appendix_derivative}
\moyenne{\phi_0(\ir\omega_n)\phi_0(\ir\omega_k)}=
\f{4\beta^2}{(k_\alpha|\omega_n|+k_C{\omega_n}^2)(k_\alpha|\omega_m|+k_C{\omega_m}^2)}\\
\times \f{1}{Z}\times \f{\delta^2Z}{\delta a(-\ir\omega_n)\,\delta a(-\ir\omega_m)}\bigg|_{a=0}
\end{multline}

\noindent Before taking the derivatives, it's convenient to perform a shift, $\phi_0(\tau)\to \phi_0(\tau)-a(\tau)/2$. The source term is eliminated and the action $S\e{tot}$ is 

\begin{multline}\label{appendix_actionsource}
S=\f{1}{2\beta}\SUM{k}{} \Big(k_\alpha\,|\omega_n|+k_C\,{\omega_n}^2\Big)
\,\bigg[|\phi_0(\ir\omega_n)|^2-\f{1}{4}\,|a(\ir\omega_n)|^2\bigg]\\
-E_J\,\displaystyle\int_{0}^{\beta}\dd\tau\,\cos\Big(\phi_0(\tau)-\f{a(\tau)}{2}\Big)
\end{multline}

\marge After taking the derivatives of Eq.\eqref{appendix_derivative}, the 
Matsubara Green's function $G(\tau)$ becomes

\begin{multline}
G(\ir\omega_n)=-\f{1}{k_\alpha\,|\omega_n|+k_C\,{\omega_n}^2}
\\
-\f{{E_J}^2}{(k_\alpha\,|\omega_n|+k_C\,{\omega_n}^2)^2}\,\displaystyle\int_{0}^{\beta}
\dd\tau\,\ex{\ir\omega_n\tau}\,\moyenne{\mathcal{T}_\tau \sin\phi_0(\tau)\sin\phi_0(0)}
\end{multline}

\noindent We want to compute the DC resistance  ($\omega\to 0$). We can 
drop the $k_C$ term in the last expression. We find 

\begin{multline}
\f{R(T)}{R\e{s}} = 1+\f{2\pi}{\alpha}\,
\limt_{\omega\to 0}\mathcal{R}\bigg[\limt_{\ir\omega_n\to \omega+\ir 0^+}\,\f{1}{|\omega_n|}\times\\
\displaystyle\int_{0}{\beta}
\dd\tau\,\ex{\ir\omega_n\tau}\,{E_J}^2\moyenne{\mathcal{T}_\tau \sin\phi_0(\tau)\,\sin\phi_0(0)}\bigg]
\end{multline}

\marge With $\phi_0=2e\,\hat{\phi}(0)$, we find 

\begin{multline}
\f{R(T)}{R\e{s}} = 1+\f{2\pi}{\alpha}\,
\limt_{\omega\to 0}\mathcal{R}\bigg[\limt_{\ir\omega_n\to \omega+\ir 0^+}\,\f{1}{|\omega_n|}\times\\
\displaystyle\int_{0}^{\beta}
\dd\tau\,\ex{\ir\omega_n\tau}\,{E_J}^2\moyenne{\mathcal{T}_\tau \sin[2e\hat{\phi}(\tau)]\,\sin[2e\hat{\phi}(0)]}\bigg]
\end{multline}

\marge When $E_M\neq 0$, the calculation is straightforward because the $E_J$-term and the $E_M$-term in Eq.\eqref{hamilto} are separated and we find Eq.\eqref{resistance} and Eq.\eqref{matsubara}.

\section{Perturbation theory for the resistance}\label{appen-perturbation}

For $E_J\ll E_C$, the hamiltonian \eqref{hamilto4} is quadratic. 
We can use the Wick's theorem and obtain the equation \eqref{correlator}:

\begin{equation}\label{wick}
\moyenne{\sin[pe\,\hat{\phi}(\tau)]\,\sin[pe\,\hat{\phi}(0)]}_0=
\f{1}{2}\,\ex{J(\tau,p)}
\end{equation}

\marge The subscript $0$ means that the average is with respect to the quadratic part of the Hamiltonian \eqref{hamilto4}. For real time and for arbitrary temperature,

\begin{equation}
J(t,p) = \f{p^2}{2}\,\displaystyle\int_{-\infty}^{+\infty}\f{\dd\omega}{\omega}\,\f{\text{Re} Z(\omega)}{R_q}\,\f{\ex{-\ir\omega t}-1}{1-\ex{-\beta \omega}}
\end{equation}

\marge In the last expression, we omit the correlator 
$\moyenne{\ex{\ir e p\hat{\phi}(\tau)}\ex{\ir e p\hat{\phi}(0)}}$ because
the exponential goes to zero. For the sake of simplicity, we set 
$E_M=0$. It doesn't change the calculation because only the combinaison between $E_J$ and itself and $E_M$ and itself give a non-zero result. 
The resistance \eqref{resistance} becomes 

\begin{equation}\label{appendixresist}
\f{R}{R\e{s}}=1+\f{\pi\,{E_J}^2}{\alpha}
\mathcal{R}\bigg(\f{\ir}{\omega}\,\limt_{\ir\omega_n\to \omega+\ir 0^+}
 \,\displaystyle\int_{0}^{\beta}\,\ex{\ir\omega_n\tau}\,
 \ex{J(\tau,2)}\bigg)
 \end{equation}
 
 \marge We can now deform the contour of integration:
 
 \begin{multline}\displaystyle\int_{0}^{\beta}
\dd\tau\,\ex{\ir\omega_n\tau}\,
\ex{J(\tau,2)}=\ir\displaystyle\int_{0}^{+\infty}\dd t\,\ex{-\omega_n t}\, \ex{J(t,2)}\\
-\ir 
\displaystyle\int_{0}^{+\infty}\dd t\,\ex{-\omega_n (t-\ir\beta)}\, \ex{J(t-\ir\beta,2)}\end{multline}

\marge With $\omega_n\beta=2\pi n$ and $\ex{J(t-\ir\beta,2)}=\ex{J(-t,2)}$, we can make the analytic continuation and find 

 \begin{multline}
 \f{R(T)}{R\e{s}}=1-\f{\pi\,{E_J}^2}{\alpha}\,\limt_{\omega\to 0}\mathcal{R}\bigg[\f{1}{\omega}\times\\
 \displaystyle\int_{0}^{+\infty}\dd t \,\ex{\ir\omega t}\,\big(\ex{J(t,2)}-\ex{J(-t,2)}\big)\bigg]\end{multline}

\marge In the DC-limit ($\omega \to 0$), 

\begin{multline}\label{pert1}
\f{R(T)}{R\e{s}}=1-\f{\pi\,{E_J}^2}{\alpha}\,\mathcal{R}\bigg[\ir \,\displaystyle\int_{0}^{+\infty}\dd t \,t\,\big(\ex{J(t,2)}-\ex{J(-t,2)}\big)\bigg]\\
 =1-\f{\pi\,{E_J}^2}{\alpha}\,\mathcal{R}\bigg[\ir \,\displaystyle\int_{-\infty}^{+\infty}\dd t \,t\,\ex{J(t,2)}\bigg]
\end{multline}

\marge Instead of integrating along the real axis, we can integration 
along the contour swept out by $t-\ir\beta/2$ for real $t$.  The last 
expression becomes 

\begin{multline}\label{perturbation3}
\f{R(T)}{R\e{s}}=1-\f{\pi\,{E_J}^2\,\beta}{2\alpha}\,\displaystyle\int_{-\infty}^{+\infty}\dd t \,\ex{J(t-\ir\beta/2,2)}\\
=1-\f{\pi\,{E_J}^2\,\beta}{\alpha}\,\displaystyle\int_{0}^{+\infty}\dd t \,
\ex{J(t-\ir\beta/2,2)}
\end{multline}

\marge where

\begin{multline}
J\Big(t-\f{\ir\beta}{2},2\Big)=j_2(t)=\f{2}{\alpha}\displaystyle\int_{0}^{+\infty}\f{\dd\omega}{\omega}\times\f{1}{1+{R\e{s}}^2C^2\omega^2}\\
\times \f{\cos(\omega t)- \text{cosh}(\beta\omega/2)}{\text{sinh}(\beta\omega/2)}\end{multline}

\marge The integration can be 
 performed exactly:

 \begin{multline}\label{integration}
j(t,2)=-\f{2}{\alpha}\,\Bigg[\f{\pi}{\beta}\,\bigg(t-\f{\ir\beta}{2}\bigg)
+\ir\pi\,\f{1-\ex{-\omega\e{RC}(t-\ir\beta/2)}}{1-\ex{\ir\beta\omega\e{RC}}}\\
+\SUM{n=1}{+\infty}\f{{\omega_{RC}}^2}{n\,({\omega_{RC}}^2-{\omega_n}^2)}\,\Big(1-\ex{-\omega_n\,(t-\ir\beta/2)}\Big)\Bigg]
\end{multline}

\marge where $\omega_n=2\pi n/\beta$ are the Matsubara frequencies and $\omega_{RC}=1/\tau\e{s}$. The sum can be expressed using special
functions:

\begin{multline}
\SUM{n=1}{+\infty}\,\f{{\omega_{RC}}^2}{n({\omega_{RC}}^2-{\omega_n}^2)}\,\Big(1-
\ex{-\omega_n(t-\ir\beta/2)}\Big)\\
=\gamma+\ln\Big(1+\ex{-2\pi t/\beta}\Big)+
\f{1}{2}\Big(\Psi(1-a)+\Psi(1+a)\Big)\\
-\f{1}{2}\,\ex{-2\pi t/\beta}\,\bigg\{
\f{1}{1-a}\,{}_2F_1\Big(1,1-a,2-a,-\ex{-2\pi t/\beta}\Big)
\\
+\f{1}{1+a}\,{}_2F_1\Big(1,1+a,2+a,-\ex{-2\pi t/\beta}\Big)\bigg\}\end{multline}

\marge where $a=\beta\omega_{RC}/(2\pi)$, $\gamma$ the Euler's constant, $\Psi$ the digamma function and ${}_2F_1$ the Hypergeometric function. Using $\Psi(1+a)=\Psi(a)+1/a$ and $\Psi(1-a)=\Psi(a)+\pi\,\cotan(\pi a)$ and making the change $y=2\pi t/\beta$ in the integral
\eqref{perturbation3}, we find 

\begin{equation}\f{R(T)}{R\e{s}}=1-\f{{E_J}^2\,\beta^2}{2\alpha}\,
\displaystyle\int_{0}^{+\infty}\dd y \,\ex{j_2(y,2\alpha E_C/\pi^2T)}\end{equation}

\marge with

\begin{multline}
j_2(y,a)=-\f{2}{\alpha}\,\Bigg\{
\gamma+\Psi(a)+\ln 2+\ln \text{ch}\paf{y}{2}+\f{1}{2}\,\bigg[\f{1}{a}+\f{\pi\,\ex{-ay}}{\sin(\pi a)}\bigg]\\
-\f{\ex{-y}}{2}\,\bigg[\f{1}{1-a}\,{}_2F_1\Big(1,1-a,2-a,-\ex{-y}\Big)
\\
+\f{1}{1+a}\,{}_2F_1\Big(1,1+a,2+a,-\ex{-y}\Big)\bigg]\Bigg\}\end{multline}

\marge This result can be extend to the case where $E_M\neq 0$ by changing $2/\alpha$ in $1/(2\alpha)$. Finally, we obtain the relation 
\eqref{perturbation}.

\section{Expression of the S-matrix}\label{appen-smatrix}

We consider a mode expansion

\begin{subequations}
\begin{align}
\hat{\psi}(x)&= \SUM{k}{} \Big(a_k(x)\,\hat{\Gamma}_k + h.c\Big)\\[3mm]
\hat{\psi}^\dagger(x)&= \SUM{k}{} \Big(b_k(x)\,\hat{\Gamma}_k + h.c\Big)\\[3mm]
\hat{a} &= \SUM{k}{} \Big({u}_k\,\hat{\Gamma}_k+ h.c\Big)
\end{align}
\end{subequations}

\marge In the
new basis, $\{\hat{\Gamma}_k,\hat{\Gamma}^\dagger_{k'}\}=\delta(k-k')$ and $\{\hat{\Gamma}_k,\hat{\Gamma}_{k'}\}=0$.  In the basis of $\hat{\Gamma}_k$, the Hamiltonian \eqref{hamiltoRF} is equal to $H=\SUM{k}{}k\,\hat{\Gamma}^\dagger_k\,\hat{\Gamma}_k$. One can obtain a Schrodinger's equation for the
wave functions $a_k(x)$, $b_k(x)$ and $u_k$ by calculating 
$[H,\hat{\psi}(x)]$, $[H,\hat{\psi}^\dagger(x)]$ and $[H,\hat{a}]$. We have

\begin{subequations}\label{system}
\begin{align}
\big[H,\hat{\psi}(x)\big] &= \ir\partial_x\hat{\psi}(x)-2\ir\,r_J\,\delta(x)\,\partial_x\hat{\psi}^\dagger(0)+r_M\,\delta(x)\,\hat{a}\\[3mm]
\big[H,\hat{\psi}^\dagger(x)\big] &= \ir\partial_x\hat{\psi}^\dagger(x)-2\ir\,r_J\,\delta(x)\,\partial_x\hat{\psi}^\dagger(0)-r_M\,\delta(x)\,\hat{a}\\[3mm]
\big[H,\hat{a}\big] &= r_M\,(\hat{\psi}(0)-\hat{\psi}^\dagger(0))
\end{align}
\end{subequations}

\marge In the new basis, the system \eqref{system} becomes

\begin{subequations}\label{system2}
\begin{align}
-k\,a_k(x)&=\ir\,\partial_xa_k-2\ir r_J\,\delta(x)\,\partial_x b_k(0)
+r_M\,\delta(x)\,u_k\\[3mm]
-k\,b_k(x) &= \ir\partial_xb_k-2\ir r_J\,\delta(x)\,\partial_x a_k(0)
- r_M\,\delta(x)\,u_k\\[3mm]
-k\,u_k &= r_M(a_k(0)-b_k(0))
\end{align}
\end{subequations}

\marge For $x\neq 0$, the solutions are 

\begin{subequations}\label{solution}
\begin{align}
a_k(x) &= \ex{\ir \epsilon_k x}\,\Big(\theta(x)\,a_k^+ +\theta(-x)\,a_k^-\Big)\\[3mm]
b_k(x) &=  \ex{\ir \epsilon_k  x}\,\Big(\theta(x)\,b_k^+ +\theta(-x)\,b_k^-\Big)
\end{align}\end{subequations}

\marge where $\theta(x)$ is the Heaviside function and 
$\epsilon_k=k$. In $x=0$, we use the following regularizations

\begin{subequations}\label{boundary}
\begin{align}
a_k(0) &= \f{a_k(0^+)+a_k(0^-)}{2}=\f{a_k^++a_k^-}{2}\\[4mm]
\partial_x a_k(0) &=\f{\partial_xa_k(0^+)+\partial_x a_k(0^-)}{2}=\f{\ir\epsilon_k}{2}\,(a_k^++a_k^-)
\end{align}\end{subequations}

\marge We have the same expression for the coefficient $b_k(x)$.
We can integrate the system \eqref{system2} around $x=0$. Using
the boundaries equations \eqref{boundary}, we can express the
coefficient $a_k^+$ with respect to $b_k^-$ and $a_k^-$:

\begin{multline}
a_k^+=b_k^-\,\f{\ir(2r_J\,{\epsilon_k}^2+{r_M}^2)}{{{r}_J}^2{\epsilon_k}^3+{r}_J{r_M}^2\,{\epsilon_k}+{\epsilon_k}+\ir{r_M}^2}
\\
-a_k^-\,
\f{-{\epsilon_k}^2+{r}_J{r_M}^2\,{\epsilon_k}+{{r}_J}^2{\epsilon_k}^3}{{{r}_J}^2{\epsilon_k}^3+{r}_J{r_M}^2\,{\epsilon_k}+{\epsilon_k}+\ir{r_M}^2}\end{multline}

\marge The particle-hole component of the S-matrix $S\e{ph}$ is
the coefficient in front of $b_k^-$. As a consequence, 

\begin{equation}
S\e{ph}(\epsilon_k)=\f{\ir(2r_J\,{\epsilon_k}^2+{r_M}^2)}{{{r}_J}^2{\epsilon_k}^3+{r}_J{r_M}^2\,{\epsilon_k}+{\epsilon_k}+\ir{r_M}^2}
\end{equation}

\end{appendix}
\newpage


\end{document}